\documentclass[twocolumn,a4paper,aps,pre,draft]{revtex4}
\usepackage{amsmath,amssymb}
\usepackage[final]{graphics}

\begin{document}

\title{Tagged-particle dynamics in a fluid adsorbed in a disordered
porous solid: interplay between the diffusion-localization and
liquid-glass transitions}

\author{V. Krakoviack}

\affiliation{Laboratoire de Chimie, UMR CNRS 5182, {\'E}cole Normale
Sup{\'e}rieure de Lyon, 46 All{\'e}e d'Italie, 69364 Lyon cedex 07,
France}

\date{\today}

\begin{abstract}
A mode-coupling theory for the slow single-particle dynamics in fluids
adsorbed in disordered porous media is derived, which complements
previous work on the collective dynamics [V. Krakoviack, Phys. Rev. E
\textbf{75}, 031503 (2007)]. Its equations, like the previous ones,
reflect the interplay between confinement-induced relaxation phenomena
and glassy dynamics through the presence of two contributions in the
slow part of the memory kernel, which are linear and quadratic in the
density correlation functions, respectively. From numerical solutions
for two simple models with pure hard core interactions, it is shown
that two different scenarios result for the diffusion-localization
transition, depending on the strength of the confinement. For weak
confinement, this transition is discontinuous and coincides with the
ideal glass transition, like in one-component bulk systems, while, for
strong confinement, it is continuous and occurs before the collective
dynamics gets nonergodic. In the latter case, the glass transition
manifests itself as a secondary transition, which can be either
continuous or discontinuous, in the already arrested single-particle
dynamics. The main features of the anomalous dynamics found in the
vicinity of all these transitions are reviewed and illustrated with
detailed computations.
\end{abstract}

\maketitle

\section{Introduction}

The dynamics of fluids confined at the nanoscale is a topic of great
interest, both for fundamental and practical reasons
\cite{Confit2000,Confit2003,Confit2006}.  Among the relevant questions
from a fundamental point of view, one counts anomalous molecular
transport characterized by subdiffusive laws, which result from the
trapping and obstruction phenomena due to geometric and topological
constraints and/or quenched disorder
\cite{HavBen02AP,BouGeo90PR,Kim02CP}. Another topic which has
attracted considerable attention in the past few years is the dynamics
of confined glass-forming liquids
\cite{AlcMcK05JPCM,AlbCoaDosDudGubRadSli06JPCM}, with the main
motivation of clarifying and substantiating the concept of
cooperativity, a key ingredient of many glass transition theories
\cite{Sil99JNCS,Edi00ARPC,Ric02JPCM}. On the practical side, the study
of many problems in the fields of biology, geology, and chemical
engineering could benefit from a finer understanding of the dynamical
behavior of confined fluids.

Recently, as a first step towards a general and versatile theoretical
approach to the problem, we have developed an extension of the ideal
mode-coupling theory (MCT) of the liquid-glass transition
\cite{BenGotSjo84JPC,Leu84PRA,LesHouches,GotSjo92RPP,Got99JPCM} which
allows one to study the slow collective dynamics of fluids adsorbed in
disordered porous solids on the basis of the model of the so-called
"quenched-annealed" (QA) binary mixture
\cite{Kra05PRL,Kra05JPCM,Kra07PRE}. In these systems, first introduced
by Madden and Glandt \cite{MadGla88JSP,Mad92JCP}, the porous solid
(the quenched component) is represented by a random array of particles
frozen in a disordered configuration sampled from a given probability
distribution, in which the fluid molecules (the annealed component)
equilibrate.

The theory makes a number of interesting predictions
\cite{Kra05PRL,Kra05JPCM,Kra07PRE}. Indeed, two distinct types of
ideal liquid-glass transition scenarios are predicted, which are
either discontinuous, like in the bulk, in situations of weak
confinement (dilute matrices), or continuous in situations of strong
confinement (dense matrices). In the intermediate region where the
nature of the transition changes, degenerate or genuine higher-order
singularities and glass-glass transition lines are found depending on
the details of the fluid-solid system. Moreover, a reentrant glass
transition line is predicted in the low fluid-high matrix density
regime.

In the present paper, we address within the same theoretical framework
the dynamics of a tagged particle moving in a fluid adsorbed in a
disordered porous solid. More precisely, we shall derive and solve the
MCT equations describing the time evolution of the
wave-vector-dependent tagged-particle density correlators and, from
their small wave vector limit, of the mean-squared displacement. These
quantities can be measured by a variety of techniques and usually more
easily than the corresponding collective properties. For instance,
because hydrogen has a large incoherent scattering length, inelastic
neutron scattering techniques essentially probe the spectrum of the
incoherent intermediate scattering function when applied to
hydrogen-rich molecular systems like water and many organic
glassformers; deuterated samples are required to measure its coherent
counterpart. Also, in computer simulations, tagged-particle quantities
can be averaged over all fluid particles in the sample, leading to a
very significant improvement in statistics compared to the collective
functions. Thus, we anticipate that with the present developments it
will become easier to compare the predictions of the theory with high
quality experimental and simulation data.

A priori, one expects to see manifestations of two types of ergodicity
breaking events in the present problem. The ideal liquid-glass
transition as obtained from the study of the collective dynamics is
the first one, following the same mechanism as in the bulk: the
tagged-particle dynamics is coupled to the collective dynamics and a
spontaneous arrest of the latter might trigger a similar arrest of the
former. The second one is the diffusion-localization transition
corresponding to the permanent trapping of the tagged particle in some
domain of finite spatial extent. This is a confinement- and
disorder-induced phenomenon which is already present when the tagged
particle is alone in the porous medium. It is one aim of the present
work to understand how these two processes interplay. Note that early
MCT studies of bulk binary hard sphere mixtures have clearly shown
that the relation between the two phenomena can be nontrivial
\cite{BosTha87PRL,ThaBos91PRA,ThaBos91PRA2}.

From the point of view of the diffusion-localization transition, the
reported theory represents an extension of early work on the classical
random Lorentz gas \cite{GotLeuYip81PRA,Leu83PRA,Sza04EL},
incorporating collective effects due to finite fluid densities at two
levels. First, through the use of the proper statistical mechanical
formalism for the definition and the computation of the static
correlation functions of the QA mixture which are needed as input of
the dynamical theory, the dependence of the structure of the fluid on
its density and on the fluid-fluid interactions is naturally accounted
for. Second, as already mentioned, the coupling of the tagged-particle
dynamics to the collective dynamics is explicitly included. So, in
addition to the localization effect of the porous medium, obstruction
phenomena due to molecular crowding or, conversely, decorrelation
processes due to fluid-fluid collisions can in principle be captured
by the theory.

The paper is organized as follows. In Sec.~II, the model of the QA
mixture is defined, the MCT equations for the collective dynamics are
recalled and those for the tagged-particle dynamics are derived and
shortly discussed. In Sec.~III, the relation of the theory to previous
work and a variant are explored, in an attempt to provide guidelines
for future studies with other formalisms. The dynamical phase diagrams
are considered in more details in Sec.~IV. Section V is devoted to the
presentation of the complete relaxation scenarios and Section VI to
concluding remarks. A few technical results are presented in
appendices.

\section{The model and its mode-coupling theory}

\begin{figure}
\includegraphics*{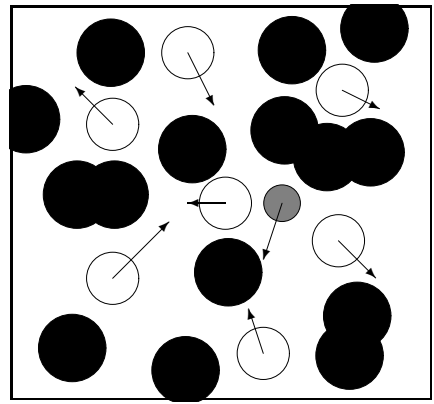}
\caption{\label{figsketch2} Sketch of a QA system. In black, the
  immobile matrix particles. In white and gray, with arrows
  symbolizing their movement, the fluid particles and the tagged
  particle, respectively.}
\end{figure}

As in MCT studies of bulk systems, the variables of interest in the
present theory are the Fourier components of the microscopic densities
characterizing the system. For the problem of the dynamics of a tagged
particle moving in a fluid adsorbed in a disordered porous solid (see
Fig.~\ref{figsketch2} for a sketch of the system), there are four such
quantities which are relevant either to the statics or the dynamics.

In a QA system, the disordered porous medium is represented by a
collection of $N_\text{m}$ rigorously immobile point particles,
randomly placed in a volume $V$ at positions denoted by $\mathbf{s}_1,
\mathbf{s}_2, \ldots, \mathbf{s}_{N_\text{m}}$ according to a given
probability distribution $\mathcal{P}(\mathbf{s}_1, \mathbf{s}_2,
\ldots, \mathbf{s}_{N_\text{m}})$ \cite{MadGla88JSP,Mad92JCP}.  Its
overall density is $n_\text{m}=N_\text{m}/V$ and its frozen density
fluctuations are given by
\begin{equation}
\rho^\text{m}_\mathbf{q}=\sum_{j=1}^{N_\text{m}} e^{i \mathbf{q}
\mathbf{s}_j},
\end{equation}
where $\mathbf{q}$ denotes the wave vector.

The fluid component consists of $N_\text{f}$ point particles (density
$n_\text{f}=N_\text{f}/V$) of mass $m$, which equilibrate at a
temperature $T$ in the random potential energy landscape created by
the frozen matrix particles. As usual, its time-dependent density
fluctuations are defined as
\begin{equation}
\rho^\text{f}_\mathbf{q}(t)=\sum_{j=1}^{N_\text{f}} e^{i \mathbf{q}
\mathbf{r}_j(t)},
\end{equation}
where $\mathbf{r}_j(t)$ is the position of the fluid particle $j$,
$j=1,\dots,N_\text{f}$, at time $t$. But, in the case of a QA system,
these quantities are not immediately useful. Indeed, because of the
random external field provided by the fixed matrix, nonzero average
density fluctuations exist at equilibrium. Thus, one is led to
consider relaxing and non-relaxing fluid density fluctuations,
corresponding to $\delta\rho^\text{f}_\mathbf{q}(t) =
\rho^\text{f}_\mathbf{q}(t) - \langle \rho^\text{f}_\mathbf{q}
\rangle$ and $\langle \rho^\text{f}_\mathbf{q} \rangle$, respectively,
where $\langle \cdots \rangle$ denotes a thermal average taken for a
given realization of the matrix.

A tagged particle of mass $m_\text{s}$ is immersed in the system. Its
time-dependent density fluctuations are given by
\begin{equation}
\rho^\text{s}_\mathbf{q}(t) = e^{i \mathbf{q} \mathbf{r}^\text{s}(t)},
\end{equation}
where $\mathbf{r}^\text{s}(t)$ is its position at time $t$. In an
infinite system, $\langle \rho^\text{s}_\mathbf{q} \rangle = 0$, so
there is no need to separate relaxing and non-relaxing parts in this
quantity.

The theory deals with disorder-averaged quantities under the
assumption that the matrix is statistically homogeneous, so that,
while for any single realization, the system lacks translational and
rotational invariance, all expectation values computed with the matrix
probability distribution have the same properties as in a truly
translationally and rotationally invariant system. For instance,
$\overline{\langle \rho^\text{f}_\mathbf{q} \rangle} = 0$, where
$\overline{\cdots}$ denotes the average over the matrix realizations
performed after the thermal average $\langle \cdots \rangle$. It also
results that the disorder-averaged correlation functions of all the
density fluctuations present in the problem are diagonal in
$\mathbf{q}$ and only depend on its modulus $q$.

The collective dynamics has been considered in
Refs.~\cite{Kra05PRL,Kra05JPCM,Kra07PRE}. It is described by a closed
set of self-consistent equations for the time evolution of the
normalized connected autocorrelation function of the fluid density
fluctuations
\begin{equation}
\phi_q(t)= \frac{\overline{ \langle \delta \rho^\text{f}_\mathbf{q}(t)
\delta \rho^\text{f}_\mathbf{-q}(0) \rangle}}{N_\text{f}
S^\text{ff(c)}_q},
\end{equation} 
where $S^\text{ff(c)}_q$ is the connected fluid-fluid structure factor
defined as
\begin{equation} \label{Sconn}
S^\text{ff(c)}_q = \frac{1}{N_\text{f}} \overline{\langle \delta
\rho^\text{f}_\mathbf{q} \delta \rho^\text{f}_\mathbf{-q} \rangle}.
\end{equation}
They consist of a standard generalized Langevin equation,
\begin{equation}\label{eqlangevin}
\ddot{\phi}_{q}(t) + \Omega_{q}^2 \phi_{q}(t) + \Omega_{q}^2 \int_0^t
d\tau \, m_q(t-\tau) \dot{\phi}_{q}(\tau)=0,
\end{equation}
with $\Omega_{q}^2= q^2 k_B T / m S^\text{ff(c)}_q$ and initial
conditions $\phi_q(0)=1$, $\dot{\phi}_q(0)=0$, and of a mode-coupling
approximation for the memory kernel, $m_q(t)=\Gamma_q \delta(t) +
m^\text{(MC)}_q(t)$, where $\Gamma_q$ is a friction coefficient
associated with fast dynamical processes and
\begin{equation} \label{kernel}
m^\text{(MC)}_q(t) = \int \frac{d^3\mathbf{k}}{(2\pi)^3} \left[
V^{(2)}_{\mathbf{q},\mathbf{k}} \phi_{k}(t) \phi_{|\mathbf{q-k}|}(t) +
V^{(1)}_{\mathbf{q},\mathbf{k}} \phi_{k}(t) \right],
\end{equation}
with
\begin{subequations} \label{vertices}
\begin{equation}\label{vtwo}
V^{(2)}_{\mathbf{q},\mathbf{k}} = \frac{1}{2} n_\text{f}
S^\text{ff(c)}_q \left[\frac{\mathbf{q}\cdot\mathbf{k}}{q^2}
\hat{c}^\text{ff(c)}_k + \frac{\mathbf{q}\cdot(\mathbf{q-k})}{q^2}
\hat{c}^\text{ff(c)}_{|\mathbf{q-k}|}\right]^2 S^\text{ff(c)}_k
S^\text{ff(c)}_{|\mathbf{q-k}|},
\end{equation}
and
\begin{equation}\label{vone}
V^{(1)}_{\mathbf{q},\mathbf{k}} = n_\text{f} S^\text{ff(c)}_q \left[
\frac{\mathbf{q}\cdot\mathbf{k}}{q^2} \hat{c}^\text{ff(c)}_k +
\frac{\mathbf{q}\cdot(\mathbf{q-k})}{q^2} \frac{1}{n_\text{f}}
\right]^2 S^\text{ff(c)}_k S^\text{ff(b)}_{|\mathbf{q-k}|}.
\end{equation}
\end{subequations}
$S^\text{ff(b)}_q$ is the disconnected or blocked fluid-fluid
structure factor
\begin{equation} \label{Sbloc}
S^\text{ff(b)}_q = \frac{1}{N_\text{f}} \overline{\langle
\rho^\text{f}_\mathbf{q} \rangle \langle \rho^\text{f}_\mathbf{-q}
\rangle},
\end{equation}
and $\hat{c}^\text{ff(c)}_q$ is the Fourier transform of the connected
fluid-fluid direct correlation function
\cite{GivSte92JCP,LomGivSteWeiLev93PRE,GivSte94PA,RosTarSte94JCP}.
Its blocked counterpart will be denoted $\hat{c}^\text{ff(b)}_q$ in
the following. For reference, the relations between the different
static correlation functions used in this work are reported in
Appendix \ref{app.oz}.  

We now turn to the tagged-particle dynamics as encoded by its density
correlator
\begin{equation}
\phi^\text{s}_q(t)= \overline{\langle \rho^\text{s}_\mathbf{q}(t)
\rho^\text{s}_\mathbf{-q}(0) \rangle}.
\end{equation} 
Using projection operator methods, a generalized Langevin equation is
first obtained,
\begin{equation}\label{eqlangevinself}
\ddot{\phi}^\text{s}_{q}(t) + \omega_{q}^2 \phi^\text{s}_{q}(t) +
\omega_{q}^2 \int_0^t d\tau \, m^\text{s}_q(t-\tau)
\dot{\phi}^\text{s}_{q}(\tau)=0,
\end{equation}
with $\omega_{q}^2= q^2 k_B T / m_\text{s}$ and initial conditions
$\phi^\text{s}_q(0)=1$, $\dot{\phi}^\text{s}_q(0)=0$. This equation is
exactly the same as for bulk systems.

It now remains to derive a mode-coupling approximation for the memory
kernel $m^\text{s}_q(t)$. Here, some care is needed. Indeed, it was
found in Ref.~\cite{Kra07PRE} that, in the case of an isolated
molecule moving in a random porous matrix, a direct calculation based
on the fact that the molecule interacts with the solid only
\cite{GotLeuYip81PRA,Leu83PRA,Sza04EL} does not yield the same result
as taking the limit of vanishing fluid density in the equations
describing the collective dynamics, while the two approaches should be
equivalent. The present problem, of which the above is actually a
special case, shows a similar difficulty. A direct approach in which
the tagged-particle density fluctuations only couple to the matrix and
fluid density fluctuations results in different equations as
considering an adsorbed binary fluid mixture in the limit of a
vanishing concentration of one of the components.  In the absence of
any obvious physical reason to consider the zero concentration case as
a special one, we make the choice of consistency at all densities and
use the latter strategy. Additional arguments in favor of this choice
and the equations obtained with the alternative approach are reported
in the next section.

So we start with the MCT equations for a mixture adsorbed in a
disordered porous matrix, which are given for completeness in appendix
\ref{MCTmixture}, specialize them to the case of a binary mixture, and
send the concentration of one component to zero. Keeping only the
leading terms, we get $m^\text{s}_q(t)=\Gamma^\text{s}_q \delta(t) +
m^\text{s(MC)}_q(t)$, where here again $\Gamma^\text{s}_q$ is a friction
coefficient associated with fast dynamical processes and
\begin{equation} \label{selfkernel}
m^\text{s(MC)}_q(t) = \int \frac{d^3\mathbf{k}}{(2\pi)^3} \left[
v^{(2)}_{\mathbf{q},\mathbf{k}} \phi^\text{s}_{k}(t) \phi_{|\mathbf{q-k}|}(t)
+ v^{(1)}_{\mathbf{q},\mathbf{k}} \phi^\text{s}_{k}(t) \right],
\end{equation}
with
\begin{subequations} \label{selfvertices}
\begin{equation}\label{selfvtwo}
v^{(2)}_{\mathbf{q},\mathbf{k}} = n_\text{f} \left[
\frac{\mathbf{q}\cdot(\mathbf{q-k})}{q^2} \right]^2 \left[
\hat{c}^\text{sf(c)}_{|\mathbf{q-k}|} \right]^2
S^\text{ff(c)}_{|\mathbf{q-k}|},
\end{equation}
and
\begin{equation}\label{selfvone}
v^{(1)}_{\mathbf{q},\mathbf{k}} = \left[
\frac{\mathbf{q}\cdot(\mathbf{q-k})}{q^2} \right]^2
\hat{h}^\text{ss(b)}_{|\mathbf{q-k}|}.
\end{equation}
\end{subequations}
$\hat{c}^\text{sf(c)}_{q}$ and $\hat{h}^\text{ss(b)}_{q}$ are the
Fourier transforms of the connected single-particle-fluid direct
correlation function and of the single-particle-single-particle
blocked total correlation function, respectively (see
Ref.~\cite{PasKah00PRE} for general definitions). They obviously have
blocked and connected counterparts, denoted by
$\hat{c}^\text{sf(b)}_{q}$ and $\hat{h}^\text{ss(c)}_{q}$,
respectively.

Like in the bulk, the last set of equations is not closed: one first
needs to solve the collective dynamics for $\phi_{q}(t)$ before one
can compute $\phi^\text{s}_{q}(t)$.  As it should, in the
$n_\text{m}\to0$ and $n_\text{f}\to0$ limits, the equations for the
tagged-particle dynamics in bulk systems \cite{LesHouches} and those
for the collective dynamics in the zero fluid density limit
\cite{Kra07PRE} are recovered, respectively. The same remarks as in
Ref.~\cite{Kra07PRE} apply: the presence of a linear term in the
memory kernel, the disappearance of any explicit reference to the
matrix, the presence of connected and disconnected structural
quantities only, and the fact that $v^{(2)}_{\mathbf{q},\mathbf{k}}$
is the same as in a bulk system, with connected quantities simply
replacing the fluid structure factor and direct correlation function.
We refer the reader to the previous work for the corresponding
discussion.

Finally, from the above formulas, one can derive an equation for the
time evolution of the mean-squared displacement $\delta
r^2(t)=\overline{\langle|\mathbf{r}^\text{s}(t) -
\mathbf{r}^\text{s}(0)|^2 \rangle}$, by using the small wave vector
behavior of the tagged-particle density correlation function,
$\phi^\text{s}_q(t)= 1 - q^2 \delta r^2(t)/6 + O(q^4)$. It reads [note
that it is possible to integrate once, thanks to the initial
conditions $\delta r^2(0)=0$ and $\dot{\delta r^2}(0)=0$]
\begin{equation}\label{eqlangevinMSD}
\dot{\delta r^2}(t) + \frac{k_B T}{m_\text{s}} \int_0^t d\tau \,
m^\text{MSD}(t-\tau) \delta r^2(\tau)=\frac{6 k_B T}{m_\text{s}}t,
\end{equation}
with $m^\text{MSD}(t) = \lim_{q\to 0} q^2 m^\text{s}_q(t) = \gamma
\delta(t) + m^\text{MSD(MC)}(t)$, where $\gamma = \lim_{q\to 0} q^2
\Gamma^\text{s}_q$ relates to the fast dynamical processes and where
the mode-coupling part of the memory kernel is given by
\begin{equation}
m^\text{MSD(MC)}(t) = \int_0^{+\infty} \frac{k^4 dk}{6 \pi^2} \left[
w^{(2)}_{k} \phi^\text{s}_{k}(t) \phi_{k}(t) + w^{(1)}_{k}
\phi^\text{s}_{k}(t) \right]
\end{equation}
with
\begin{equation}\label{MSDvertices}
w^{(2)}_{k} = n_\text{f} \left[\hat{c}^\text{sf(c)}_{k} \right]^2
S^\text{ff(c)}_{k} \text{ and } w^{(1)}_{k} =
\hat{h}^\text{ss(b)}_{k}.
\end{equation}

\section{Connections with previous theories and a variant}

The physics of QA mixtures involves a number of subtleties, related,
for instance, to the splitting of the static correlations into
connected and disconnected parts. These might induce special
difficulties and put constraints on the proper way to set up a
dynamical theory for these systems. In this technical section, we
illustrate how they manifest themselves within the MCT framework, in
an attempt to provide guidelines for future work based on other
formalisms (see Ref.~\cite{ChaJuaMed08PRE}, for instance).

In Ref.~\cite{Kra07PRE}, the MCT equations for the collective dynamics
in a QA binary mixture have been compared to those for the residual
dynamics of a bulk system in its ideal glassy phase. Strong analogies
have been found in the overall structure of both theories, which
originate in the presence of time-persistent density fluctuations in
both cases. But there are significant and irreducible differences as
well, which reflect the different origins of these frozen
fluctuations, static and disorder-induced in the former case,
dynamical and self-induced in the latter.

The main conclusion from this study was that, in general, it does not
seem sensible to try and derive a dynamical theory for QA systems by
simply taking the limit of the corresponding approach for fully
annealed mixtures in which one component representing the solid matrix
would become immobile, or at least extreme care should be taken. Since
this contradicts naive expectations, we begin the present section with
a similar analysis of the MCT for the tagged-particle dynamics. This
is clearly a simpler problem, since by construction no collective
dynamical phenomena come into play and there is no disconnected
component in the tagged-particle density correlation function.

In order to complete the discussion of Ref.~\cite{Kra07PRE}, we first
consider the case of a one-component bulk fluid in its ideal glassy
phase. The tagged-particle dynamics as expressed in terms of the
residual relaxation of the fluid density fluctuations obeys
mode-coupling equations of the same form as above, with vertices
\cite{LesHouches}
\begin{subequations} \label{selfverticesresidual}
\begin{equation}
v^{(2)}_{\mathbf{q},\mathbf{k}} = n_\text{f} \left[
\frac{\mathbf{q}\cdot(\mathbf{q-k})}{q^2} \right]^2 \left[
\hat{c}^\text{sf}_{|\mathbf{q-k}|} \right]^2 \left\{(1-f_{|\mathbf{q-k}|})
S^\text{ff}_{|\mathbf{q-k}|}\right\},
\end{equation}
and
\begin{equation}
v^{(1)}_{\mathbf{q},\mathbf{k}} = n_\text{f} \left[
\frac{\mathbf{q}\cdot(\mathbf{q-k})}{q^2} \right]^2 \left[
\hat{c}^\text{sf}_{|\mathbf{q-k}|} \right]^2 \left\{f_{|\mathbf{q-k}|}
S^\text{ff}_{|\mathbf{q-k}|}\right\},
\end{equation}
\end{subequations}
where $n_\text{f}$ is the density of the fluid, $S^\text{ff}_q$ its
structure factor, $\hat{c}^\text{sf}_q$ the Fourier transform of the
single-particle-fluid direct correlation function, and $f_q$ the
Debye-Waller factor of the glass. To make contact with the theory for
QA systems, we now need an expression for $\hat{h}^\text{ss(b)}_{q}$
in Eq.~\eqref{selfvone}. Since the present problem corresponds to a
situation where there are fluid density fluctuations only, we shall
take advantage of the fact that the matrix does not appear explicitly
in the MCT equations for QA systems and use the Ornstein-Zernike (OZ)
equations for a simpler example of a fluid in a random environment,
namely, a fluid plunged in a Gaussian random field \cite{whatOZ}.  One
then gets \cite{MenDas94PRL,KieRosTar99JSP}
\begin{equation} \label{hssb1}
\hat{h}^\text{ss(b)}_{q} = \hat{c}^\text{ss(b)}_{q} + 2 n_\text{f}
\hat{c}^\text{sf(b)}_{q} \hat{c}^\text{sf(c)}_{q} S^\text{ff(c)}_{q} +
n_\text{f} \left[ \hat{c}^\text{sf(c)}_{q} \right]^2
S^\text{ff(b)}_{q},
\end{equation}
where $\hat{c}^\text{ss(b)}_{q}$ is the Fourier transform of the
single-particle-single-particle blocked direct correlation function.
Thus, we observe that there is an exact correspondence between the two
theories, with the same role played by $S^\text{ff(c)}_{q}$ and
$(1-f_q) S^\text{ff}_q$ on the one hand, $S^\text{ff(b)}_{q}$ and $f_q
S^\text{ff}_q$ on the other hand, provided one disregards in
Eq.~\eqref{hssb1} the terms involving blocked direct correlation
functions. This in particular insures that the full and connected
direct correlation functions coincide.

As a second step, let us now approach the problem following the
empirical strategy outlined above, i.e., we shall consider the QA
binary mixture as a special limit of a fully annealed binary
mixture. As already pointed out, this approach is ineffective for the
description of the collective dynamics \cite{Kra07PRE}. For a tagged
particle moving in a fully annealed binary mixture, the MCT result for
the memory kernel reads (for simplicity, we keep the same species
labels as in the QA binary mixture)
\cite{Got87NATO,BosTha87PRL,ThaBos91PRA2}
\begin{multline}
m^\text{s(MC)}_q(t) = (n_\text{f}+n_\text{m}) \int
 \frac{d^3\mathbf{k}}{(2\pi)^3} \left[
 \frac{\mathbf{q}\cdot(\mathbf{q-k})}{q^2} \right]^2
 \phi^\text{s}_k(t) \\
\left\{ \left[\hat{c}^\text{sf}_{|\mathbf{q-k}|}\right]^2
F^\text{ff}_{|\mathbf{q-k}|}(t) + 2 \hat{c}^\text{sf}_{|\mathbf{q-k}|}
\hat{c}^\text{sm}_{|\mathbf{q-k}|} F^\text{fm}_{|\mathbf{q-k}|}(t)
\right. \\
\left. + \left[\hat{c}^\text{sm}_{|\mathbf{q-k}|}\right]^2
 F^\text{mm}_{|\mathbf{q-k}|}(t) \right\},
\end{multline}
where $\hat{c}^\text{sm}_q$ is the Fourier transform of the
single-particle-matrix direct correlation function and, in the
definitions of the density correlation functions $F^\text{ff}_q(t)$,
$F^\text{fm}_q(t)$, and $F^\text{mm}_q(t)$, the usual normalization
for binary mixtures has been used, hence the prefactor
$n_\text{f}+n_\text{m}$. We might now adapt this equation to the case
of a QA binary mixture, for which $F^\text{ff}_q(t)$ splits into
relaxing and frozen parts according to
\begin{subequations}
\begin{equation}
(n_\text{f}+n_\text{m}) F^\text{ff}_q(t) \equiv n_\text{f}
S^\text{ff(b)}_{q} + n_\text{f} S^\text{ff(c)}_{q} \phi_q(t),
\end{equation}
while $F^\text{fm}_q(t)$ and $F^\text{mm}_q(t)$ are actually
time-independent and given by
\begin{gather}
(n_\text{f}+n_\text{m}) F^\text{fm}_q(t) \equiv \sqrt{n_\text{f}
n_\text{m}} S^\text{fm}_q, \\ (n_\text{f}+n_\text{m}) F^\text{mm}_q(t)
\equiv n_\text{m} S^\text{mm}_q.
\end{gather} 
\end{subequations}
$S^\text{fm}_q$ and $S^\text{mm}_q$ are the fluid-matrix and
matrix-matrix structure factors, respectively, and, like
$S^\text{ff(b)}_{q}$ and $S^\text{ff(c)}_{q}$, they are defined with
the normalization suitable for QA mixtures (see Appendix
\ref{app.oz}). One then obtains mode-coupling equations of the same
form as above, with vertices
\begin{subequations} \label{selfverticesempirical}
\begin{equation}
v^{(2)}_{\mathbf{q},\mathbf{k}} = n_\text{f} \left[
\frac{\mathbf{q}\cdot(\mathbf{q-k})}{q^2} \right]^2 \left[
\hat{c}^\text{sf}_{|\mathbf{q-k}|} \right]^2
S^\text{ff(c)}_{|\mathbf{q-k}|} ,
\end{equation}
and
\begin{multline}
v^{(1)}_{\mathbf{q},\mathbf{k}} = \left[
\frac{\mathbf{q}\cdot(\mathbf{q-k})}{q^2} \right]^2 \left( n_\text{f}
\left[\hat{c}^\text{sf}_{|\mathbf{q-k}|}\right]^2
S^\text{ff(b)}_{|\mathbf{q-k}|} \right. \\ \left. + 2 \sqrt{n_\text{f}
n_\text{m}} \hat{c}^\text{sf}_{|\mathbf{q-k}|}
\hat{c}^\text{sm}_{|\mathbf{q-k}|} S^\text{fm}_{|\mathbf{q-k}|} +
n_\text{m} \left[\hat{c}^\text{sm}_{|\mathbf{q-k}|}\right]^2
S^\text{mm}_{|\mathbf{q-k}|} \right).
\end{multline}
\end{subequations}
If this result is confronted to the relevant expression of
$\hat{h}^\text{ss(b)}_{q}$ \cite{PasKah00PRE},
\begin{multline}\label{hssb2}
\hat{h}^\text{ss(b)}_{q} = \hat{c}^\text{ss(b)}_{q} + 2 n_\text{f}
\hat{c}^\text{sf(b)}_{q} \hat{c}^\text{sf(c)}_{q} S^\text{ff(c)}_{q} +
n_\text{f} \left[ \hat{c}^\text{sf(c)}_q \right]^2 S^\text{ff(b)}_{q}
\\ + 2 \sqrt{n_\text{f} n_\text{m}} \hat{c}^\text{sf(c)}_q
\hat{c}^\text{sm}_q S^\text{fm}_q + n_\text{m}
\left[\hat{c}^\text{sm}_q\right]^2 S^\text{mm}_q,
\end{multline}
an exact correspondence with equations \eqref{selfvertices} is once
again found, under the same conditions as in the previous calculation.

So, we find that, in both cases, there would be a perfect consistency
between the derived equations and the theory developed for fluids in
random environments if all blocked direct correlation functions were
identically zero. Or, stated differently, both approaches would be
perfectly acceptable starting points for heuristic derivations of the
present theory if this condition was fulfilled. Obviously, it is not,
and this provides further arguments in favor of a differentiated
treatment of fully annealed and quenched-annealed mixtures.

However, at the same time, the approximation which consists in
neglecting the blocked direct correlation functions is a very common
one, which for historical reasons is often referred to as the
Madden-Glandt approximation
\cite{GivSte92JCP,LomGivSteWeiLev93PRE,GivSte94PA}. Its widespread use
originates in the fact that these functions are among the specific
features of fluids in quenched disordered environments which are the
most difficult to capture with simple approximations, and, for
instance, many standard closures of the replica OZ equations
erroneously prescribe that they vanish identically
\cite{GivSte92JCP,LomGivSteWeiLev93PRE,GivSte94PA,MerLevWei96JCP}.  It
is thus hardly surprising that they appear in the errors made when
trying to infer a theory for QA systems starting from a similar
approach for fully annealed systems, for which these functions are
meaningless.

The crucial simplification in the case of the tagged-particle
dynamics, compared to the collective dynamics, is that the
Madden-Glandt approximation seems to be the \emph{only} required
approximation \cite{collective}. This is explicitly demonstrated here
in the framework of the MCT, but we expect that analogous situations
could occur with other theoretical schemes. So, the conclusion could
be that it does not appear unreasonable to try and derive empirical
theories for the tagged-particle dynamics in QA systems starting from
approaches developed for fully equilibrated systems, but this will
usually be at the cost of additional implicit approximations like
Madden-Glandt's. Fortunately, most of the time, such approximations
have quantitative consequences only and the cost appears modest. Note
however that one serious restriction remains: since no such simple
empirical scheme seems to be generically applicable to the collective
dynamics \cite{Kra07PRE}, it will always be difficult to give a proper
account of the situations in which this part of the dynamics matters,
for instance, if the coupling of the single-particle and collective
dynamics is strong.

Another illustration of the subtle interplay between the dynamical
theory and the peculiar structure of the static correlations in fluids
evolving in random environments is provided by the alternative
derivation of the MCT for the tagged-particle dynamics in QA systems
which has been discarded in the previous section. It is based on the
observation that the only forces exerted on the tagged molecule are
those due to the fluid and the random matrix. So, on simple physical
grounds, one expects that the only mode-coupling contributions to the
relaxation kernel will come from the products
$\rho^\text{s}_\mathbf{k} \delta \rho^\text{f}_\mathbf{q-k}$,
$\rho^\text{s}_\mathbf{k} \langle \rho^\text{f}_\mathbf{q-k} \rangle$,
and $ \rho^\text{s}_\mathbf{k} \rho^\text{m}_\mathbf{q-k}$. Projecting
the random forces on this set of variables using standard projection
operator methods, one obtains a memory kernel which is essentially the
same as above, except that one has to replace in Eq.~\eqref{selfvone}
$\hat{h}^\text{ss(b)}_{|\mathbf{q-k}|}$ as given by Eq.~\eqref{hssb2}
by $\hat{h}^\text{ss(b)}_{|\mathbf{q-k}|} -
\hat{c}^\text{ss(b)}_{|\mathbf{q-k}|} + \left[
\hat{c}^\text{sf(b)}_{|\mathbf{q-k}|} \right]^2 /
\hat{c}^\text{ff(b)}_{|\mathbf{q-k}|}$.

An immediate reason to reject this approach is that, in the limit
$n_\text{f}\to0$ where the dynamics should be determined by the
properties of the matrix and of the tagged particle only, an
unphysical dependence on the fluid would remain in the theory because
of the last term of this expression. This is avoided with the
derivation retained in the previous section, which is thus clearly
preferable.

The only difference between the two approaches is that, by first
working with a finite concentration of tagged particles, the correct
one incorporates contributions associated with their frozen density
fluctuations, which seem to be preserved when the concentration is
sent to zero. In this respect, it is pictorial that the two approaches
happen to agree when the fluid and tagged particles are identical,
thanks to the equalities $\hat{c}^\text{ss(b)}_{q} =
\hat{c}^\text{sf(b)}_{q} = \hat{c}^\text{ff(b)}_{q}$. Indeed, it is in
this unique case that the frozen fluid and tagged-particle density
fluctuations share the same statistics and that the former can act in
the incorrect approach as substitutes for the latter which would
otherwise be missing. Thus, the inclusion of contributions due to the
frozen tagged-particle density fluctuations, which are rather elusive
as far as species with vanishing concentration are concerned, appears
as an important requirement in order to derive an acceptable theory.
Retrospectively, failure to take into account similar contributions in
special limiting cases (vanishing fluid density, adsorbed ideal gas)
of the collective dynamics surely explains some ambiguous results
pointed out in Ref.~\cite{Kra07PRE}.  At present, the physics, if any,
behind this constraint is unclear.

\section{Dynamical phase diagrams}

We now turn to the quantitative results of the theory, which require
numerical solutions of the MCT equations. In the present section, we
complete the dynamical phase diagrams of Ref.~\cite{Kra07PRE}, where
the liquid-glass transition lines for two simple model systems had
been determined, while, in the next one, the full dynamical scenarios
for the tagged-particle motion will be discussed.

The systems to be considered are two closely related QA mixture models
in which both the fluid-fluid and fluid-matrix interactions are pure
hard core repulsions of the same diameter $d$. The only difference
between them lies in the matrix correlations. In model I, the matrix
particles are not allowed to overlap, so that the configurational
statistics of the porous medium is that of an equilibrium hard sphere
fluid, while in model II the matrix particles are completely
uncorrelated and overlap freely, as in an ideal gas. For both systems,
the two dimensionless densities $\phi_\text{f}=\pi n_\text{f} d^3/6$
and $\phi_\text{m}=\pi n_\text{m} d^3/6$ will be used as control
parameters, and the structural quantities required by the dynamical
theory will be computed with the Percus-Yevick (PY) approximation
\cite{GivSte92JCP,LomGivSteWeiLev93PRE,GivSte94PA,MerLevWei96JCP}.

For simplicity, we shall concentrate on the case where the tagged
particle is one of the fluid particles, so that in
Eqs.~\eqref{selfvertices} and \eqref{MSDvertices} one has to take
$\hat{c}^\text{sf(c)}_{q}= \hat{c}^\text{ff(c)}_{q}$ and
$\hat{h}^\text{ss(b)}_{q}= \hat{h}^\text{ff(b)}_{q}$. Then, the
tagged-particle dynamics coincides with the self part of the
collective dynamics.

From the point of view of the single-particle dynamics, two distinct
phases exist in MCT. If for all $q$ $\lim_{t\to\infty}
\phi^\text{s}_q(t)=0$, one is in the diffusive regime, where the
tagged particle can explore an infinite domain and move infinitely far
away from its initial position. If on the contrary for all $q$
$\lim_{t\to\infty} \phi^\text{s}_q(t)=f^\text{s}_q\neq0$, one is in
the localized regime, where the tagged particle is trapped in a finite
domain.

The single-particle nonergodicity parameter $f^\text{s}_q$, which is
usually called the Lamb-M{\"o}ssbauer factor, is the solution of the
nonlinear set of equations
\begin{equation}\label{nonlinear}
\frac{f^\text{s}_q}{1-f^\text{s}_q}=\int
\frac{d^3\mathbf{k}}{(2\pi)^3} \left[ v^{(2)}_{\mathbf{q},\mathbf{k}}
f^\text{s}_{k} f_{|\mathbf{q-k}|} + v^{(1)}_{\mathbf{q},\mathbf{k}}
f^\text{s}_{k} \right],
\end{equation}
where $f_{q}$ is the collective dynamics nonergodicity parameter or
Debye-Waller factor computed in Ref.~\cite{Kra07PRE}. This equation
has to be solved numerically in order to locate the diffusive and
localized phases when $\phi_\text{f}$ and $\phi_\text{m}$ are
varied. All computations in the present paper have been achieved using
the methods of Refs.~\cite{FraFucGotMaySin97PRE,FucGotMay98PRE}, to
which the interested reader is referred for technical details, with
the same parameters as in Ref.~\cite{Kra07PRE}. As pointed out in the
latter work, the discretization used for the wave vector integration
in Eq.~\eqref{nonlinear} involves a cutoff of the low $q$ divergence
of the memory kernels. This has the advantage of eliminating spurious
long time anomalies originating in this divergence
\cite{Got81PMB,Leu83PRAb,HofFreFraUNPUB}, but at the cost of
quantitative uncertainties due to the arbitrariness of the cutoff
value. They have been estimated in two ways, first by test
calculations on a finer $q$ grid, second by a comparison with the
results obtained within an additional hydrodynamic approximation (see
Appendix \ref{hydro}) which allows one to integrate exactly over the
full $q$ range \cite{Leu83PRA}. It has been found that the errors on
the location of the transition points never exceed a few percents.

\begin{figure*}
\includegraphics*{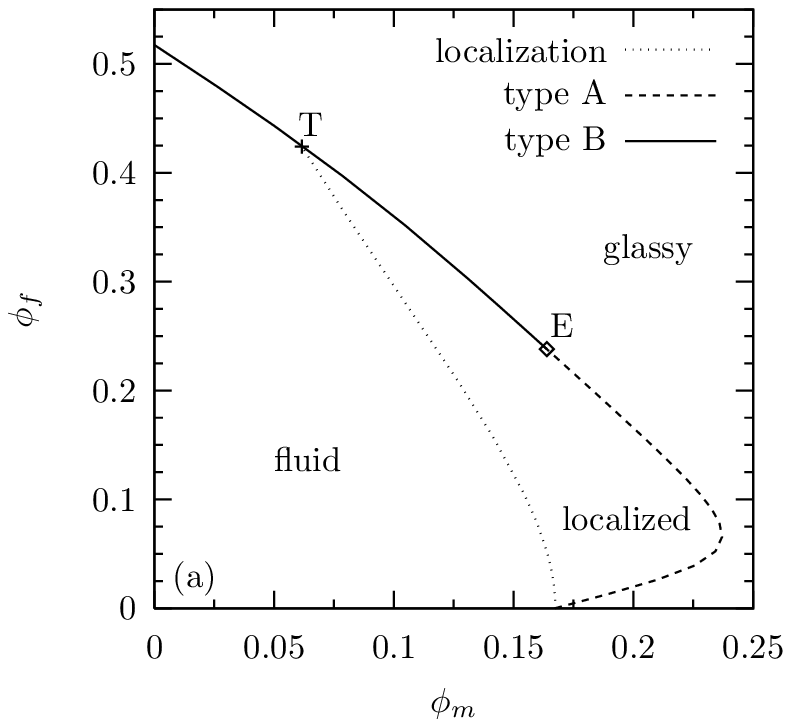}\hfill\includegraphics*{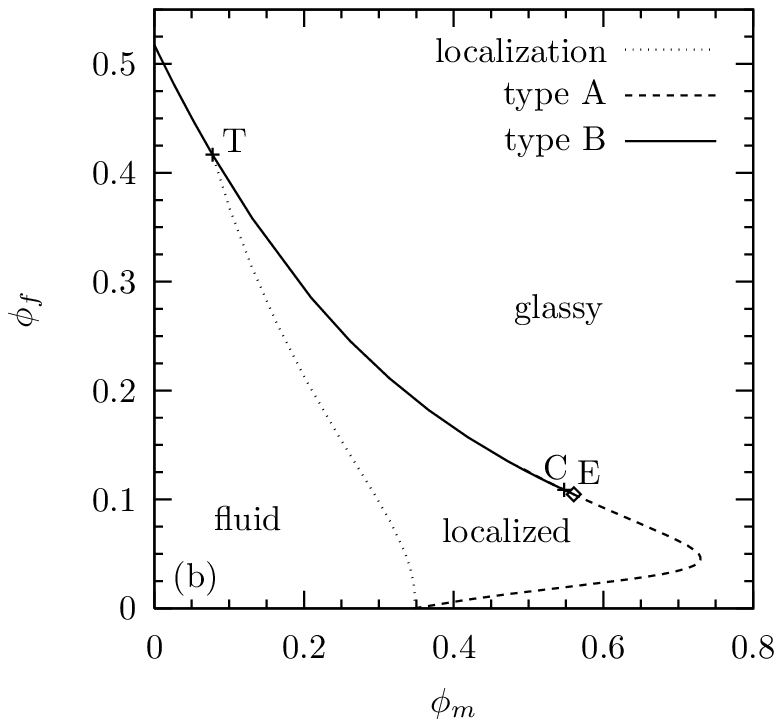}
\caption{\label{figtrans} Dynamical phase diagrams of two types of
  hard-sphere quenched-annealed binary mixtures. (a) Model I:
  non-overlapping matrix particles. (b) Model II: freely overlapping
  matrix particles. For the collective dynamics, the $A_3$ endpoint
  singularities are denoted by E and the crossing point between the
  type A and B transition lines in model II is denoted by C. For the
  tagged-particle dynamics, the crossing points between the
  diffusion-localization and glass transition lines are denoted by T
  (for triple point).}
\end{figure*}

The dynamical phase diagrams of models I and II are reported in
Fig.~\ref{figtrans}. The liquid-glass transition lines of types A and
B are those obtained in Ref.~\cite{Kra07PRE}. As discussed there, for
model I they join smoothly at a degenerate higher-order $A_3$
singularity, while for model II a crossing point followed by a line of
glass-glass transitions terminating at an ordinary $A_3$ singularity
is obtained. Except for these details, both phase diagrams are very
similar at the level of the collective dynamics.

This remains true when the tagged-particle dynamics is considered. For
both models, two types of transitions between the diffusive and
localized phases are found. 

At low matrix densities, the glass and localization transitions occur
simultaneously. In fact, it is the spontaneous arrest of the
collective dynamics which triggers the arrest of the tagged-particle
dynamics through their coupling in the bilinear term of the memory
kernel \eqref{selfkernel}. This is the well known scenario found in
all MCT studies of one-component bulk systems so far, in which the
tagged-particle dynamics inherits its discontinuous (type B) character
and quantitative features (critical laws, exponent parameter
$\lambda$) from the collective dynamics
\cite{LesHouches,FucGotMay98PRE}. Its extension to weakly confined
systems is easily understood from simple continuity arguments in the
vicinity of the bulk transition at $\phi_\text{m}=0$.

These arguments break down at higher matrix densities, where one finds
that the tagged particles are already localized when the ideal glass
transition occurs. Thus, the latter only manifests itself as a
secondary dynamical transition in an already arrested single-particle
dynamics, with characteristic features imposed by the collective
dynamics just like in the previous scenario. In fact, in this density
domain, the liquid-glass transition is always preceded by a continuous
diffusion-localization transition driven by the linear term of the
memory kernel \eqref{selfkernel}. These transitions, which only
manifest themselves in the single-particle dynamics, are special cases
of type A transitions, characterized by an exponent parameter
$\lambda=0$, like the pure diffusion-localization transition at
$\phi_\text{f}=0$ \cite{GotLeuYip81PRA,Leu83PRA,Kra07PRE}. Similar
ones have been previously discussed in the framework of schematic
models \cite{LesHouches,Sjo86PRA,GotHau88ZPB,FraGot94JPCM}, and, more
interestingly, have also been found in early MCT studies of bulk
binary hard sphere mixtures with large size asymmetry, where they take
place in the ideal glassy phase and correspond to the localization of
the smaller species in the voids of the glass
\cite{BosTha87PRL,ThaBos91PRA,ThaBos91PRA2}. Superficially, the
analogy between the two systems can be understood by seeing the
disordered porous solid in the QA mixture as a glassy component or by
considering the induction of permanent fluid density fluctuations by
the quenched random environment as some kind of previtrification.  But
we stress once again that, as illustrated in Ref.~\cite{Kra07PRE} and
in the previous section, such simple pictures do not in general lend
themselves to the development of rigorous approaches to the dynamics
of QA systems.

As seen in Fig.~\ref{figtrans}, the corresponding transition line
starts on the $\phi_\text{f}=0$ axis at the same point as the
liquid-glass transition line, since in this limit the collective
dynamics exactly reduces to the single-particle dynamics. As
$\phi_\text{f}$ is increased, it runs across the fluid domain as
obtained from the study of the collective dynamics, and it terminates
when it intercepts the type B liquid-glass transition line at some
finite value of $\phi_\text{m}$.

The final result is thus that the dynamical phase diagrams of both QA
mixture models display three domains, as shown in
Fig.~\ref{figtrans}. For low overall densities, the dynamics is
fluid-like, with both the collective and tagged-particle density
fluctuations relaxing to their equilibrium values at long times. For
intermediate densities, the dynamics is localized, with a complete
relaxation of the collective density fluctuations only. And for high
densities, glassy dynamics is found, in which neither the collective
nor the single-particle density fluctuations do return to
equilibrium. These three domains meet at two points: first, at the
pure diffusion-localization transition point at $\phi_\text{f}=0$,
because, as mentioned above, there coincide the collective and
tagged-particle dynamics; second, at the crossing point between the
liquid-glass and continuous diffusion-localization transition
lines. For definiteness, because of the three phases, we shall call
this point a triple point denoted by T in Fig.~\ref{figtrans}. But it
should be remembered that phase coexistence is excluded in MCT,
because of a maximum property of the physical solution of the MCT
equations \cite{LesHouches}. It results that point T corresponds
unambiguously to a glassy state.

The clear separation between the diffusion-localization and
liquid-glass transition lines allows one to completely disregard the
collective dynamics when dealing with the asymptotic properties of the
tagged-particle dynamics in the vicinity of the former
transition. This possibility is particularly interesting if an
additional hydrodynamic approximation is performed, since significant
analytic progress can then be made in the study of the continuous
diffusion-localization transition.  This approach, which extends
Leutheusser's theory \cite{Leu83PRA} to finite fluid densities, is
discussed in Appendix \ref{hydro}.

\begin{figure}
\includegraphics*{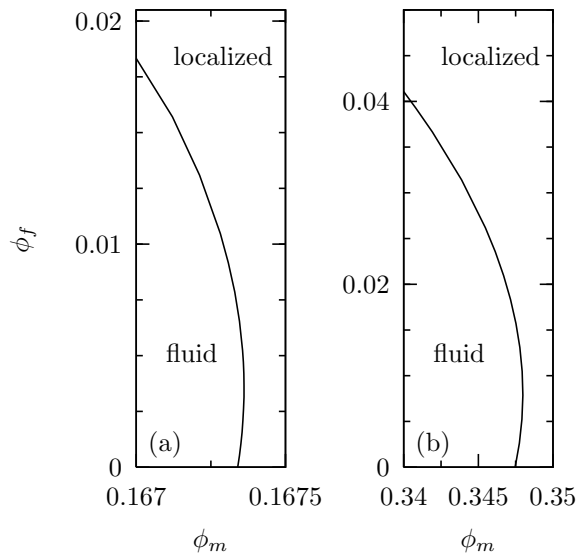}
\caption{\label{figtranszoom} Magnified views of the continuous
  diffusion-localization transition lines for high matrix
  densities. A reentrant behavior is clearly visible. (a) Model I,
  (b) Model II.}
\end{figure}

Beside the nature of the domains and transition lines met in the phase
diagrams, their shape is of interest as well. A remarkable result of
Ref.~\cite{Kra07PRE} has been the prediction of a reentry phenomenon
in the liquid-glass transition line for high matrix-low fluid
densities, which has been interpreted as the signature of a dynamical
decorrelation mechanism due to fluid-fluid collisions competing with
the reduction of free volume and the localization effect due to the
disordered porous solid. A similar feature, which might be understood
using the same physical arguments, is also present in the continuous
diffusion-localization transition line, but with a much weaker
amplitude, so that one needs the strong magnifications of
Fig.~\ref{figtranszoom} in order to visualize it.

As already pointed out in Ref.~\cite{Kra07PRE}, a word of caution is
needed about this finding. Indeed, for systems ruled by hard core
fluid-matrix interactions like those studied in the present work, a
number of recent extensive simulation studies \cite{HofFraFre06PRL,%
HofFra07PRL,HofMunFreFra08JCP,SunYet06PRL,SunYet08JPCB,%
SunYet08JCP,BabGimNic08JPCB} have clearly demonstrated that the
localization transition is actually driven by the percolation
transition of the matrix void space, i.e., localization occurs
because, above a certain critical matrix density, the void space only
consists of finite disconnected domains. Such a scenario definitely
rules out the possibility of a reentry phenomenon in the
diffusion-localization transition line as obtained above. Indeed, if
at a given value of $\phi_\text{m}$ a system with $\phi_\text{f}=0$ is
localized because of the onset of the percolation transition, any
system with a finite $\phi_\text{f}$ at the same matrix density will
be localized as well, since a variation of $\phi_\text{f}$ obviously
has no effect on the topology of the matrix empty space.

Fortunately, the amplitude of the reentry phenomenon predicted by the
MCT for the diffusion-localization transition is very small, so the
inconsistency is quantitatively not too serious, but, clearly, this
prediction should not be taken literally. Instead, a reasonable
expectation is typically a nonmonotonic variation of the
self-diffusion coefficient in the slowly diffusive regime just below
the localization threshold, with first an increase due to fluid-fluid
collisions when increasing $\phi_\text{f}$ from zero.  In support of
this suggestion, we recall that, in a number of systems for which the
MCT has predicted reentrant transition lines, such a reentry has been
observed in the isodiffusivity curves as obtained by molecular
dynamics simulations \cite{FofDawBulSciZacTar02PRE,%
ZacFofDawBulSciTar02PRE,FofSciTarZacLovReaDawLik03PRL,%
ZacLowWesSciTarLik04PRL,MayZacStiLikSciMunGauHadIatTarLowVla08NM,%
ChoMorSciKob05PRL,MorChoKobSci05JCP}. The same behavior should also be
expected in systems with soft repulsive fluid-matrix interactions, for
which the percolation concepts are only approximate. In this respect,
it is encouraging that this is precisely what has been observed in a
computer simulation study of a two-dimensional lattice gas model with
fixed randomly placed penetrable scatterers \cite{Ole91JPA}.

\section{Dynamical scenarios}

According to the phase diagrams reported in the previous section, both
QA mixtures considered in this work display three generic types of
global dynamical scenarios in response to increases of the overall
density of the fluid-matrix system. For small $\phi_\text{m}$, a
bulk-like type B ideal glass transition scenario prevails, in which
both the collective and single-particle density fluctuations
simultaneously show discontinuities. For moderate $\phi_\text{m}$, a
continuous diffusion-localization transition first occurs, followed at
a higher density by a type B ideal glass transition. Finally, for high
$\phi_\text{m}$, the latter transition is of type A.

It is the purpose of this section to provide illustrations of these
three typical scenarios and to discuss their features which might be
relevant for comparisons of experimental or simulation data with the
predictions of the theory. As in Ref.~\cite{Kra07PRE}, the analytic
asymptotic results shall be quoted without their proofs, which can be
found in or adapted from Ref.~\cite{LesHouches} (see also
Ref.~\cite{FraGot94JPCM} for type A transitions). Note that these
results, which are obtained under the assumption of regular vertices,
as well as the method of solution of the mode-coupling equations used
in this work \cite{FraFucGotMaySin97PRE,FucGotMay98PRE}, which
involves a cutoff of the memory kernels at low $q$, cannot give an
account of possible dynamical features originating in the divergence
of the mode-coupling vertices when $q\to0$
\cite{Got81PMB,Leu83PRAb,HofFreFraUNPUB}. While this might look
unsatisfactory from a mathematical point of view, such an
approximation is in fact required for physical reasons, since the low
$q$ singularity is actually known to be an ill feature of the
mode-coupling approximation which results in spurious long time
anomalies \cite{Leu83PRAb}.

For simplicity and consistency with Ref.~\cite{Kra07PRE}, we follow
Refs.~\cite{FraFucGotMaySin97PRE} and \cite{FucGotMay98PRE} and
consider the dynamics in the overdamped limit valid for Brownian
systems. The generalized Langevin equations \eqref{eqlangevinself} and
\eqref{eqlangevinMSD} then reduce to
\begin{equation}
\tau^\text{s}_{q} \dot{\phi}^\text{s}_{q}(t) + \phi^\text{s}_{q}(t) +
\int_0^t d\tau \, m^\text{s(MC)}_q(t-\tau)
\dot{\phi}^\text{s}_{q}(\tau) = 0
\end{equation}
and 
\begin{equation}
\delta r^2(t) + D_0 \int_0^t d\tau \, m^\text{MSD(MC)}(t-\tau) \delta
r^2(\tau) = 6 D_0 t,
\end{equation}
with $D_0$ the short-time diffusivity, $\tau^\text{s}_{q} = 1 / (D_0
q^2)$, and the initial conditions $\phi^\text{s}_q(0)=1$ and $\delta
r^2(0)=0$. This simplification changes the short time transient part
of the dynamics, but not its long time properties. In the following,
the particle diameter $d$ is used as the unit of length and the unit
of time is chosen such that $D_0=1/160$. Also, since both QA models
are found essentially equivalent, only model I will be considered.

For the bulk-like scenario found for small $\phi_\text{m}$, it is now
well known from extensive studies of one-component bulk systems
\cite{LesHouches,FucGotMay98PRE} that, in this case, the essential
features of the tagged-particle dynamics are simple reflections of
those of the collective dynamics, since it is the latter which
actually drives the system through its unique ergodicity breaking
transition. So we do not repeat the description of the type B scenario
given in Ref.~\cite{Kra07PRE}, which remains valid in the present case
without change, and we omit the analogues for $f^\text{s}_q$ and
$\phi^\text{s}_{q}(t)$ of Figs.~5(a), 7(a), and 8(a) of
Ref.~\cite{Kra07PRE}, in which the evolutions with density of $f_q$
and $\phi_{q}(t)$ were reported, since they would display exactly the
same patterns. Instead, we concentrate on the distinctive aspects of
the single-particle dynamics, i.e., the wave vector dependence of the
Lamb-M{\"o}ssbauer factor and the time dependence of the mean-squared
displacement. As in Ref.~\cite{Kra07PRE}, we choose
$\phi_\text{m}=0.05$ for the present illustrations.

\begin{figure}
\includegraphics*{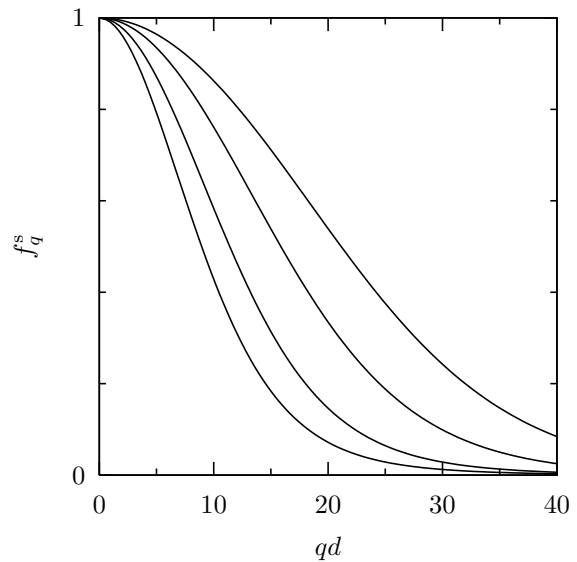}
\caption{\label{figscen1fvsq} Wave vector dependence of the
  single-particle nonergodicity parameter or Lamb-M{\"o}ssbauer factor
  $f^\text{s}_q=\lim_{t\to\infty} \phi^\text{s}_q(t)$ for model I at
  matrix density $\phi_\text{m}=0.05$. From left to right, bottom to
  top: $\phi_\text{f}=\phi_\text{f}^\text{c}$,
  $1.01\phi_\text{f}^\text{c}$, $1.05\phi_\text{f}^\text{c}$, and
  $1.1\phi_\text{f}^\text{c}$. }
\end{figure}

Fig.~\ref{figscen1fvsq} shows $f^\text{s}_q$ as a function of $q$ at
different fluid densities in the glassy phase. As in bulk systems, the
curves start at 1 for $q=0$, are approximately Gaussian-shaped and
widen rapidly when $\phi_\text{f}$ is increased. For a given distance
to the transition, they are generically narrower than the
corresponding curves for the bulk, reflecting the fact that, when
$\phi_\text{m}$ increases, the system evolves towards a continuous
glass transition scenario.

\begin{figure}
\includegraphics*{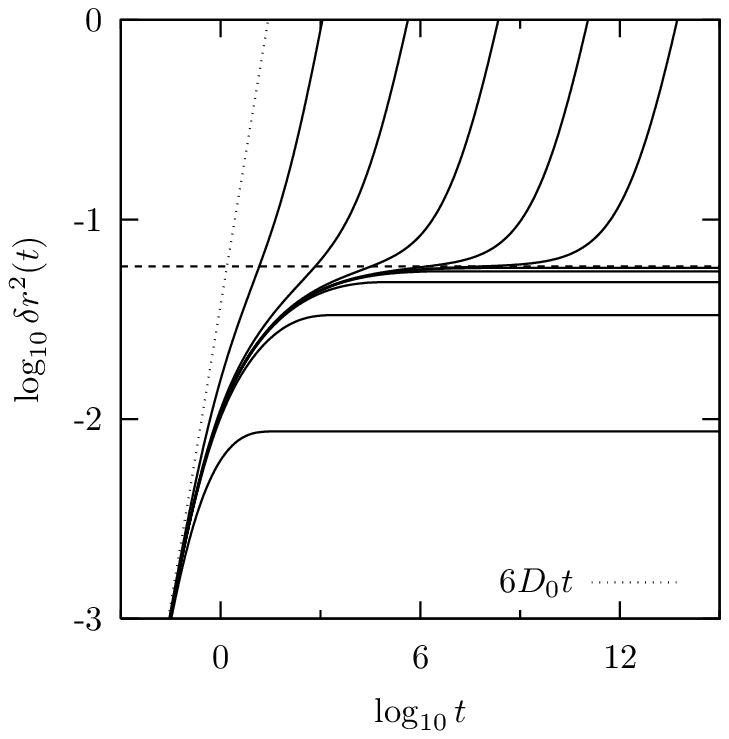}
\caption{\label{figscen1msd} Time evolution of the mean-squared
displacement $\delta r^2(t)$ for model I at matrix density
$\phi_\text{m}=0.05$, in the vicinity of the liquid-glass transition
threshold $\phi_\text{f}^\text{c}$. From left to right, top to bottom:
$\phi_\text{f} = 0.9\phi_\text{f}^\text{c}$,
$0.99\phi_\text{f}^\text{c}$, $0.999\phi_\text{f}^\text{c}$,
$0.9999\phi_\text{f}^\text{c}$, $0.99999\phi_\text{f}^\text{c}$,
$1.00001\phi_\text{f}^\text{c}$, $1.0001\phi_\text{f}^\text{c}$,
$1.001\phi_\text{f}^\text{c}$, $1.01\phi_\text{f}^\text{c}$,
$1.1\phi_\text{f}^\text{c}$.  The horizontal dashed line marks $6
r_\text{l}^\text{c2}$, with $r_\text{l}^\text{c}$ the critical
localization length. The dotted line shows the short-time diffusive
behavior $\delta r^2(t) = 6 D_0 t$.}
\end{figure}

The mean-squared displacements at different fluid densities near the
critical one are reported in Fig.~\ref{figscen1msd}. For small $t$,
short-time diffusion with diffusivity $D_0$ is observed. Then, in the
ergodic phase, the familiar two-step dynamics typical of the
discontinuous ideal glass transition scenario sets in, with, as
$\phi_\text{f}^\text{c}$ is approached, the development of a plateau
reflecting transient localization, which crosses over for large $t$ to
the diffusion behavior $\delta r^2(t)= 6 D t$. The long-time diffusion
coefficient is generically given by \cite{FucGotMay98PRE,notediff}
\begin{equation}
D= \frac{D_0}{\displaystyle 1+D_0 \int_0^\infty m^\text{MSD(MC)}(t)
dt} .
\end{equation} 
and thus vanishes in the present scenario according to the $\alpha$
relaxation scaling law, i.e., $\propto |\phi_\text{f} -
\phi_\text{f}^\text{c}|^{1/2a+1/2b}$, where $a$ and $b$ represent the
critical decay and von Schweidler exponents of the collective
dynamics, respectively. When the transition density is reached and
exceeded, the plateau lasts forever, the tagged particle being
permanently trapped. Close to the bifurcation point, the localization
length $r_\text{l}$, defined through
\begin{equation}
\lim_{t\to\infty} \delta r^2(t) = 6 r_\text{l}^2,
\end{equation}
so that
\begin{equation}
r_\text{l}^2=\left[\int_0^{+\infty} \frac{k^4 dk}{6
\pi^2} \left( w^{(2)}_{k} f^\text{s}_{k} f_{k} + w^{(1)}_{k}
f^\text{s}_{k} \right) \right]^{-1},
\end{equation}
displays a square-root singularity, $r_\text{l} - r_\text{l}^\text{c}
\propto - (\phi_f-\phi^c_f)^{1/2}$, with $r_\text{l}^\text{c}$ the
finite localization length at the transition. In both the diffusive
and localized phases, the dynamics in the vicinity of the plateau is
described by the same scaling functions as the density fluctuations,
with a critical amplitude $\propto |\phi_\text{f} -
\phi_\text{f}^\text{c}|^{1/2}$ and a characteristic time scale
$\propto |\phi_\text{f} - \phi_\text{f}^\text{c}|^{-1/2a}$.

The two other scenarios, occuring for moderate and high
$\phi_\text{m}$, have in common to start with a continuous
diffusion-localization transition which affects only the
tagged-particle dynamics and leaves the collective quantities free of
any singularity. It will now be discussed in some detail and
illustrated with computations at $\phi_\text{m}=0.1$.

\begin{figure}
\includegraphics*{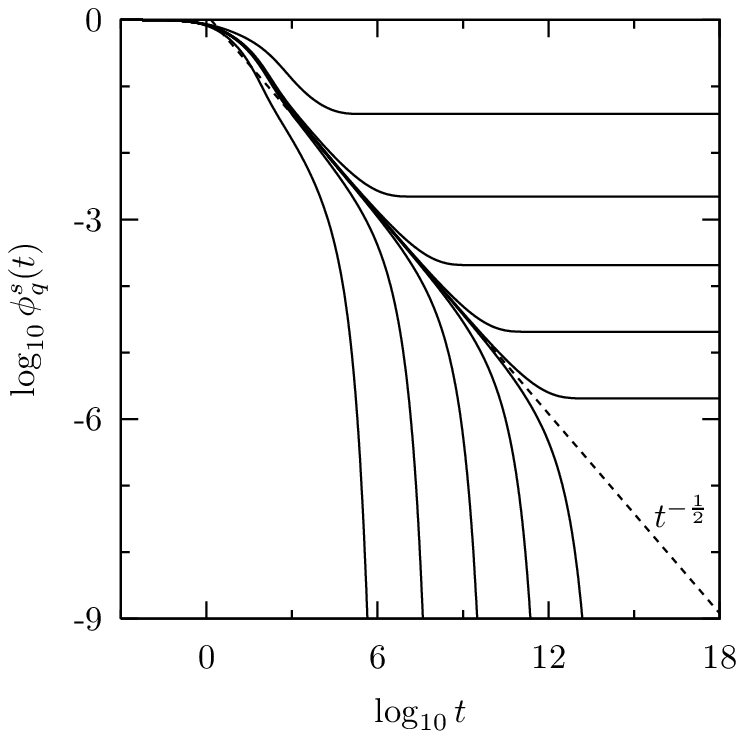}
\caption{\label{figlocaldynscal} Time evolution of the single-particle
  density correlation function $\phi^\text{s}_{q}(t)$ at $q \simeq
  7.09/d$ for model I at matrix density $\phi_\text{m}=0.1$, in the
  vicinity of the localization threshold $\phi_\text{f}^\text{l}$.
  From left to right, bottom to top: $\phi_\text{f} =
  0.9\phi_\text{f}^\text{l}$, $0.99\phi_\text{f}^\text{l}$,
  $0.999\phi_\text{f}^\text{l}$, $0.9999\phi_\text{f}^\text{l}$,
  $0.99999\phi_\text{f}^\text{l}$, $1.00001\phi_\text{f}^\text{l}$,
  $1.0001\phi_\text{f}^\text{l}$, $1.001\phi_\text{f}^\text{l}$,
  $1.01\phi_\text{f}^\text{l}$, $1.1\phi_\text{f}^\text{l}$.  The
  dashed line shows the analytically derived critical decay law.}
\end{figure}

The time evolution of the single-particle density correlation function
$\phi^\text{s}_{q}(t)$ at $q \simeq 7.09/d$, corresponding to the main
peak of $S^\text{ff(c)}_q$, is reported in Fig.~\ref{figlocaldynscal}
in a log-log plot (the curves for other values of $q$ are
qualitatively similar). In both the diffusive and localized phases
near the transition, a single step relaxation is found.  When the
localization threshold $\phi_\text{f}^\text{l}$ is approached from
below, the slowing-down of the dynamics manifests itself through the
development of a weak long time tail which extends to longer and
longer times as the density is increased. At the transition, it lasts
indefinitely, and, above the threshold, it progressively recedes and
continuously turns into a finite Lamb-M{\"o}ssbauer factor.  From a
quantitative point of view, the MCT diffusion-localization transition
is most easily understood as a type A bifurcation characterized by an
exponent parameter $\lambda=0$ and thus a critical decay exponent
$a=1/2$. It results that, close to the transition in the localized
phase, $f^\text{s}_q \propto (\phi_\text{f} -
\phi_\text{f}^\text{l})$. A reduction theorem also holds, stating
that, in both phases and for long enough times, the wave vector and
time dependence of $\phi^\text{s}_{q}(t)$ factorize according to
\begin{equation} \label{factor}
\phi^\text{s}_{q}(t) - \lim_{t\to\infty} \phi^\text{s}_{q}(t) =
h^\text{s}_q G^\text{s}(t).
\end{equation}
At the threshold, one gets the critical decay
\begin{equation} \label{critical}
G^\text{s}(t)=(t^\text{s}_0/t)^{1/2},
\end{equation}
where $t^\text{s}_0$ is a time scale obtained by matching the short
and long time dynamics, while, for finite values of
$|\phi_\text{f}-\phi_\text{f}^\text{l}|$, a scaling law is found,
\begin{equation}
G^\text{s}(t) = c^\text{s} g^\text{s}(t/\tau^\text{s}),
\end{equation}
with $c^\text{s} \propto |\phi_\text{f}-\phi_\text{f}^\text{l}|$ and
$\tau^\text{s}/t^\text{s}_0 \propto |\phi_\text{f} -
\phi_\text{f}^\text{l}|^{-2}$. $g^\text{s}(\hat{t})$ is discussed in
detail in Ref.~\cite{LesHouches} and one finds in particular that
\begin{equation} \label{abovefc}
g^\text{s}(\hat{t}) = \hat{t}^{-1/2}
\end{equation}
for small $\hat{t}$, so that the critical decay law is reached
continuously when $\phi_\text{f}\to\phi_\text{f}^\text{l}$. Note that
the sign of $\phi_\text{f}-\phi_\text{f}^\text{l}$ is irrelevant in
the dynamical scaling laws. An important feature of all type A
scenarios follows, which is the symmetric departure from the critical
decay law \eqref{critical} at long times for state points located in
the ergodic and nonergodic phases at the same distance from the
transition. All these analytic predictions are easily confirmed by
inspection of Fig.~\ref{figlocaldynscal}.

\begin{figure}
\includegraphics*{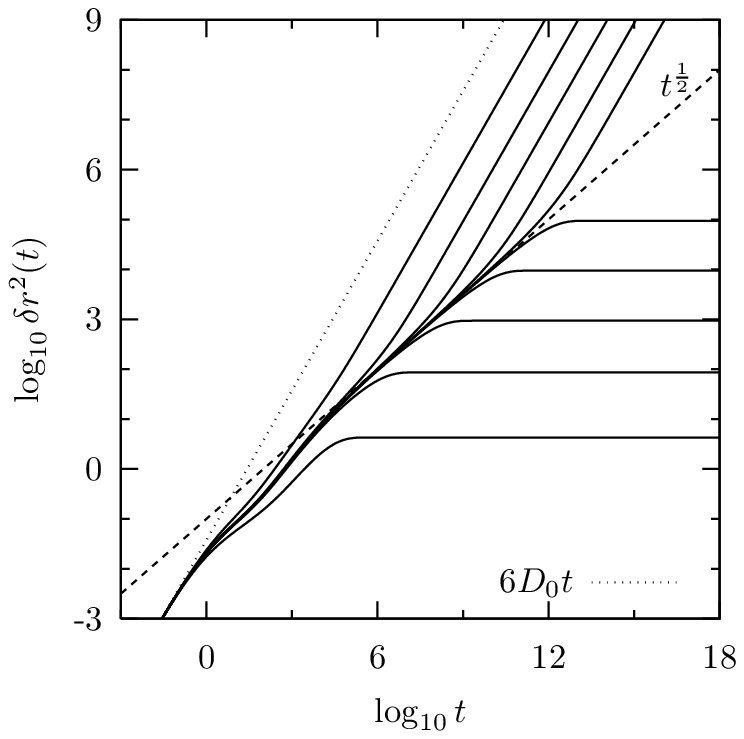}
\caption{\label{figlocalmsd} Time evolution of the mean-squared
displacement $\delta r^2(t)$ for model I at matrix density
$\phi_\text{m}=0.1$, in the vicinity of the localization threshold
$\phi_\text{f}^\text{l}$. From left to right, top to bottom:
$\phi_\text{f} = 0.9\phi_\text{f}^\text{l}$,
$0.99\phi_\text{f}^\text{l}$, $0.999\phi_\text{f}^\text{l}$,
$0.9999\phi_\text{f}^\text{l}$, $0.99999\phi_\text{f}^\text{l}$,
$1.00001\phi_\text{f}^\text{l}$, $1.0001\phi_\text{f}^\text{l}$,
$1.001\phi_\text{f}^\text{l}$, $1.01\phi_\text{f}^\text{l}$,
$1.1\phi_\text{f}^\text{l}$.  The dotted and dashed lines show,
respectively, the short-time diffusive behavior $\delta r^2(t) = 6 D_0
t$ and the analytically derived critical subdiffusive law $\delta
r^2(t)\propto t^{1/2}$.}
\end{figure}

Since, according to Eqs.~\eqref{eqlangevinMSD}--\eqref{MSDvertices},
the time evolution of the mean-squared displacement is controled by
that of the single-particle density fluctuations, the anomalous
dynamics of the latter near the diffusion-localization transition is
naturally transfered to the former. Indeed, as seen in
Fig.~\ref{figlocalmsd}, the long time tail of $\phi^\text{s}_{q}(t)$
induces the development in $\delta r^2(t)$ of a subdiffusive regime at
intermediate times, where $\delta r^2(t) \propto t^{1/2}$. This regime
would persist forever exactly at the transition. For finite
$\phi_\text{f}-\phi_\text{f}^\text{l}$, it ends at a time of the order
of $\tau^\text{s}$, where it crosses over to ordinary diffusion
$\delta r^2(t)= 6 D t$ in the diffusive phase or to a constant $\delta
r^2(t) \simeq 6 r_\text{l}^2$ in the localized phase. When the
transition density is approached from below, the long-time diffusion
coefficient vanishes linearly, $D \propto |\phi_\text{f} -
\phi_\text{f}^\text{c}|$, while the localization length diverges
according to $r_\text{l} \propto |\phi_\text{f} -
\phi_\text{f}^\text{c}|^{-1/2}$ when the threshold is approached from
above. Here again, all these asymptotic laws can be easily checked in
the figure.

Once localization has occured, further increase of the overall density
drives the system towards its ideal liquid-glass transition, which can
be either discontinuous or continuous.  The first case is realized for
$\phi_\text{m}=0.1$, which was considered above from the point of view
of the diffusion-localization transition. To illustrate the second
case, we will choose $\phi_\text{f}=0.15$ and use $\phi_\text{m}$ as
the external control parameter, since, with the QA mixture models
studied in this work, it is not possible to find a suitable constant
$\phi_\text{m}$ trajectory in the dynamical phase diagrams. Note that
this exchange of the roles played by $\phi_\text{f}$ and
$\phi_\text{m}$ does not affect the general critical properties of the
theory, provided they are formulated in terms of the relevant control
parameter. These critical properties will not be discussed in
detail. Indeed, it can be very easily demonstrated, by introducing
shifted and rescaled tagged-particle density correlation functions so
as to eliminate the contributions due to localization, that there is
actually no formal difference between the present situation and the
simpler case of the bulk-like scenario. So, here as well, the
anomalous dynamics of the tagged-particle density fluctuations in the
vicinity of the ideal liquid-glass transition is completely controled
by the collective dynamics, which imposes the main qualitative and
quantitative features of the relaxation.

\begin{figure}
\includegraphics*{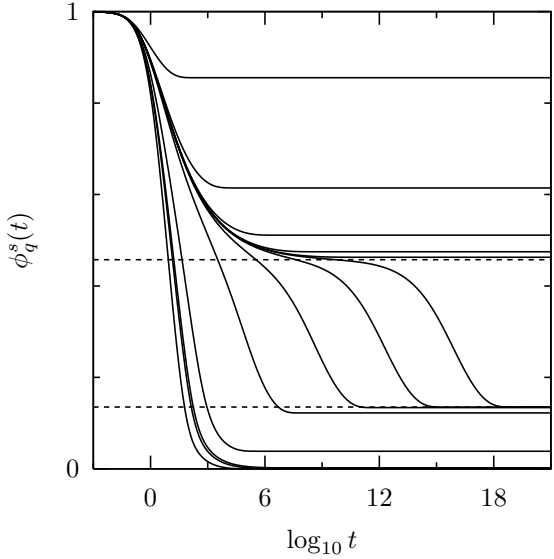}
\caption{\label{figscen2dyn} Time evolution of the single-particle
  density correlation function $\phi^\text{s}_{q}(t)$ at $q \simeq
  7.09/d$ for model I at matrix density $\phi_\text{m}=0.1$, in the
  vicinity of the diffusion-localization and liquid-glass transition
  thresholds $\phi_\text{f}^\text{l}$ and $\phi_\text{f}^\text{g}$.
  From left to right, bottom to top: $\phi_\text{f} =
  0.9\phi_\text{f}^\text{l}$, $0.99\phi_\text{f}^\text{l}$,
  $1.01\phi_\text{f}^\text{l}$, $1.1\phi_\text{f}^\text{l} \simeq
  0.9\phi_\text{f}^\text{g}$, $0.99\phi_\text{f}^\text{g}$,
  $0.999\phi_\text{f}^\text{g}$, $0.9999\phi_\text{f}^\text{g}$,
  $0.99999\phi_\text{f}^\text{g}$, $1.00001\phi_\text{f}^\text{g}$,
  $1.0001\phi_\text{f}^\text{g}$, $1.001\phi_\text{f}^\text{g}$,
  $1.01\phi_\text{f}^\text{g}$, $1.1\phi_\text{f}^\text{g}$. The lower
  and higher horizontal dashed lines mark the Lamb-M{\"o}ssbauer
  factors in the limits $\phi_\text{f}\to{\phi_\text{f}^\text{g}}^-$
  and $\phi_\text{f}\to{\phi_\text{f}^\text{g}}^+$, respectively.}
\end{figure}

\begin{figure}
\includegraphics*{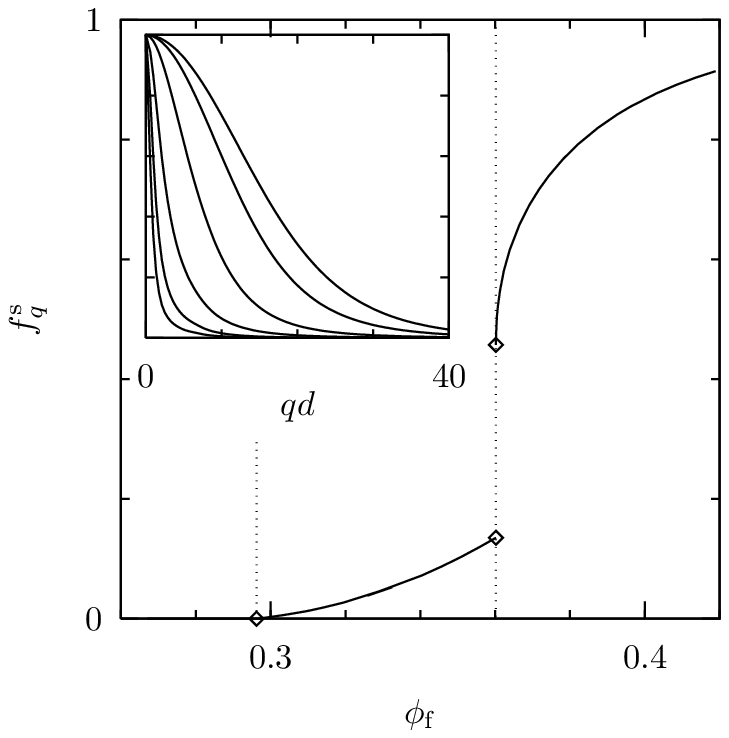}
\caption{\label{figscen2fs} Density dependence of the single-particle
  nonergodicity parameter or Lamb-M{\"o}ssbauer factor
  $f^\text{s}_q=\lim_{t\to\infty} \phi^\text{s}_q(t)$ at $q \simeq
  7.09/d$ for model I at matrix density $\phi_\text{m}=0.1$. Vertical
  dotted lines and diamonds indicate the diffusion-localization and
  liquid-glass transitions.  Inset: Wave vector dependence of
  $f^\text{s}_q$. From left to right, bottom to top:
  $\phi_\text{f}=1.05 \phi_\text{f}^\text{l}$,
  $1.1\phi_\text{f}^\text{l}$, ${\phi_\text{f}^\text{g}}^-$,
  ${\phi_\text{f}^\text{g}}^+$, $1.05 \phi_\text{f}^\text{g}$, and
  $1.1\phi_\text{f}^\text{g}$.}
\end{figure}

\begin{figure}
\includegraphics*{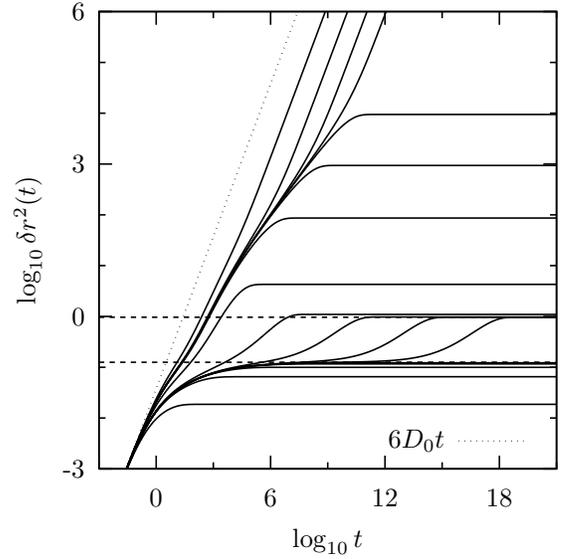}
\caption{\label{figscen2msd} Time evolution of the mean-squared
displacement $\delta r^2(t)$ for model I at matrix density
$\phi_\text{m}=0.1$, in the vicinity of the diffusion-localization and
liquid-glass transition thresholds $\phi_\text{f}^\text{l}$ and
$\phi_\text{f}^\text{g}$.  From left to right, top to bottom:
$\phi_\text{f} = 0.9\phi_\text{f}^\text{l}$,
$0.99\phi_\text{f}^\text{l}$, $0.999\phi_\text{f}^\text{l}$,
$0.9999\phi_\text{f}^\text{l}$, $1.0001\phi_\text{f}^\text{l}$,
$1.001\phi_\text{f}^\text{l}$, $1.01\phi_\text{f}^\text{l}$,
$1.1\phi_\text{f}^\text{l} \simeq 0.9\phi_\text{f}^\text{g}$,
$0.99\phi_\text{f}^\text{g}$, $0.999\phi_\text{f}^\text{g}$,
$0.9999\phi_\text{f}^\text{g}$, $0.99999\phi_\text{f}^\text{g}$,
$1.00001\phi_\text{f}^\text{g}$, $1.0001\phi_\text{f}^\text{g}$,
$1.001\phi_\text{f}^\text{g}$, $1.01\phi_\text{f}^\text{g}$,
$1.1\phi_\text{f}^\text{g}$. The higher and lower horizontal dashed
lines mark $6 r_\text{l}^\text{2}$ in the limits
$\phi_\text{f}\to{\phi_\text{f}^\text{g}}^-$ and
$\phi_\text{f}\to{\phi_\text{f}^\text{g}}^+$, respectively.  The
dotted line shows the short-time diffusive behavior $\delta r^2(t) = 6
D_0 t$.}
\end{figure}

The complete double transition scenario at $\phi_\text{m}=0.1$ is
illustrated in Figs.~\ref{figscen2dyn}, \ref{figscen2fs}, and
\ref{figscen2msd}. The data corresponding to the lowest densities
provide further illustrations of the features mentioned above for the
dynamics near the diffusion-localization transition threshold
$\phi_\text{f}^\text{l}$. This includes the continuous development of
the long time tail and finite asymptote in $\phi^\text{s}_{q}(t)$, now
shown in a semi-log plot in Fig.~\ref{figscen2dyn}, and the continuous
and linear increase of $f^\text{s}_q$ from zero at
$\phi_\text{f}^\text{l}$, as seen in Fig.~\ref{figscen2fs}.  Then,
approaching the glass transition density $\phi_\text{f}^\text{g}$, the
single-particle density correlation functions start to display a two
step decay, which is clearly visible in Fig.~\ref{figscen2dyn} and is
naturally expected from a type B glass transition scenario. When the
second relaxation step disappears because of the spontaneous arrest of
the collective dynamics, a discontinuity followed by a square root
singularity occurs in the Lamb-M{\"o}ssbauer factor, as seen in
Fig.~\ref{figscen2fs}, whose inset shows that $f^\text{s}_q$ as a
function of $q$ always remains bell-shaped. All these anomalous
behaviors are transmitted to the mean-squared displacement through the
memory kernel $m^\text{MSD(MC)}(t)$ and result in the two step
relaxation pattern visible in Fig.~\ref{figscen2msd}.

\begin{figure}
\includegraphics*{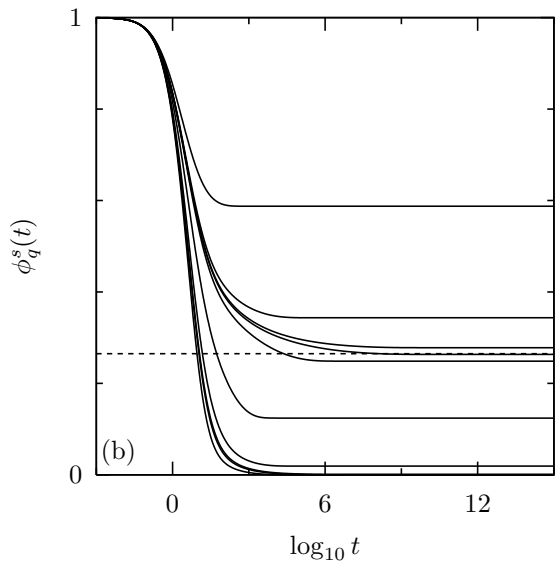}
\caption{\label{figscen3dyn} Time evolution of the single-particle
  density correlation function $\phi^\text{s}_{q}(t)$ at $q \simeq
  7.09/d$ for model I at fluid density $\phi_\text{f}=0.15$, in the
  vicinity of the diffusion-localization and liquid-glass transition
  thresholds $\phi_\text{m}^\text{l}$ and $\phi_\text{m}^\text{g}$.
  From left to right, bottom to top: $\phi_\text{m} =
  0.9\phi_\text{m}^\text{l}$, $0.99\phi_\text{m}^\text{l}$,
  $1.01\phi_\text{m}^\text{l}$, $1.1\phi_\text{m}^\text{l}$,
  $0.9\phi_\text{m}^\text{g}$, $0.99\phi_\text{m}^\text{g}$,
  $0.999\phi_\text{m}^\text{g}$, $1.001\phi_\text{m}^\text{g}$,
  $1.01\phi_\text{m}^\text{g}$, $1.1\phi_\text{m}^\text{g}$. The
  horizontal dashed line marks the Lamb-M{\"o}ssbauer factor at
  $\phi_\text{m}^\text{g}$.}
\end{figure}

\begin{figure}
\includegraphics*{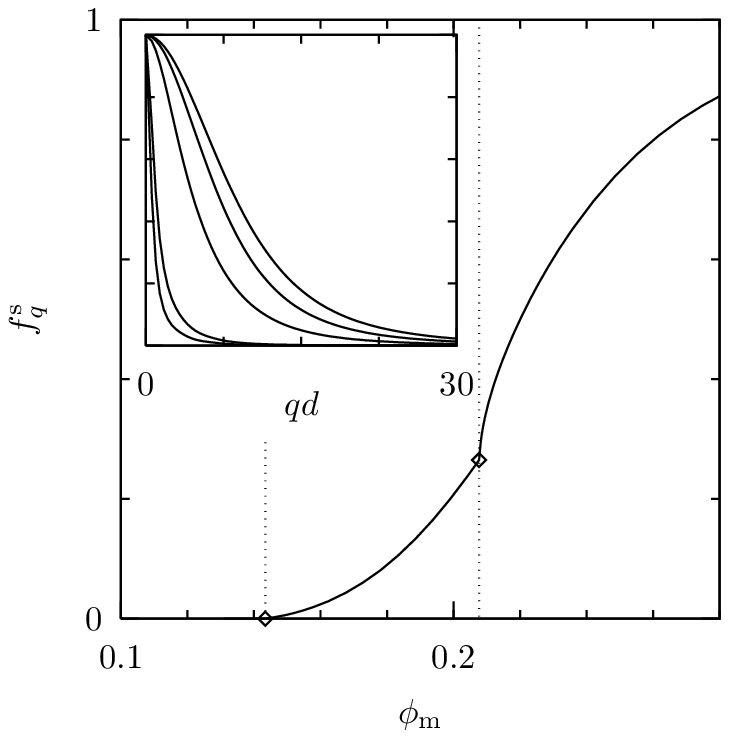}
\caption{\label{figscen3fs} Density dependence of the single-particle
  nonergodicity parameter or Lamb-M{\"o}ssbauer factor
  $f^\text{s}_q=\lim_{t\to\infty} \phi^\text{s}_q(t)$ at $q \simeq
  7.09/d$ for model I at fluid density $\phi_\text{f}=0.15$. Vertical
  dotted lines and diamonds indicate the diffusion-localization and
  liquid-glass transitions.  Inset: Wave vector dependence of
  $f^\text{s}_q$. From left to right, bottom to top:
  $\phi_\text{m}=1.05 \phi_\text{m}^\text{l}$,
  $1.1\phi_\text{m}^\text{l}$, $\phi_\text{m}^\text{g}$, $1.05
  \phi_\text{m}^\text{g}$, and $1.1\phi_\text{m}^\text{g}$.  }
\end{figure}

\begin{figure}
\includegraphics*{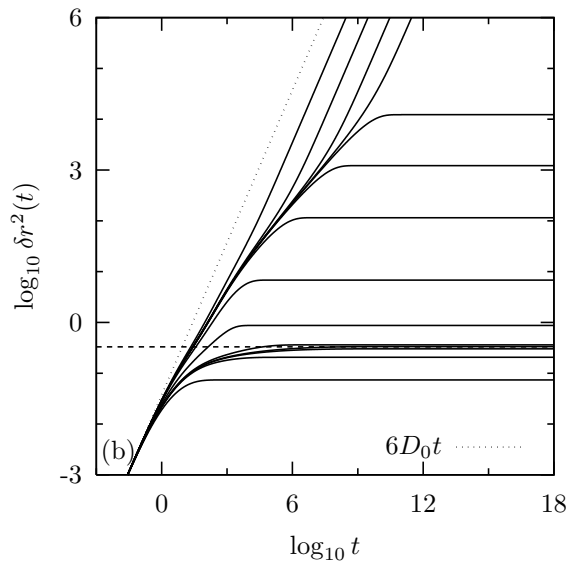}
\caption{\label{figscen3msd} Time evolution of the mean-squared
displacement $\delta r^2(t)$ for model I at fluid density
$\phi_\text{f}=0.15$, in the vicinity of the diffusion-localization
and liquid-glass transition thresholds $\phi_\text{m}^\text{l}$ and
$\phi_\text{m}^\text{g}$.  From left to right, top to bottom:
$\phi_\text{m} = 0.9\phi_\text{m}^\text{l}$,
$0.99\phi_\text{m}^\text{l}$, $0.999\phi_\text{m}^\text{l}$,
$0.9999\phi_\text{m}^\text{l}$, $1.0001\phi_\text{m}^\text{l}$,
$1.001\phi_\text{m}^\text{l}$, $1.01\phi_\text{m}^\text{l}$,
$1.1\phi_\text{m}^\text{l}$, $0.9\phi_\text{m}^\text{g}$,
$0.99\phi_\text{m}^\text{g}$, $0.999\phi_\text{m}^\text{g}$,
$1.001\phi_\text{m}^\text{g}$, $1.01\phi_\text{m}^\text{g}$,
$1.1\phi_\text{m}^\text{g}$. The horizontal dashed line marks $6
r_\text{l}^\text{2}$ at $\phi_\text{m}^\text{g}$. The dotted line
shows the short-time diffusive behavior $\delta r^2(t) = 6 D_0 t$.}
\end{figure}

The analogous data for the double transition scenario at
$\phi_\text{f}=0.15$ are reported in Figs.~\ref{figscen3dyn},
\ref{figscen3fs}, and \ref{figscen3msd}. Compared to the previous
case, there is no significant difference from the point of view of the
dynamics near the diffusion-localization transition threshold
$\phi_\text{m}^\text{l}$. But, as the matrix density is increased, the
relaxation of the tagged-particle density fluctuations always remains
single stepped, as shown by Fig.~\ref{figscen3dyn}. The type A
liquid-glass transition signals itself by a considerable slowing down
of the approach by $\phi^\text{s}_{q}(t)$ of its infinite time value,
which procedes via an algebraic decay law with the same exponent as
the collective dynamics. The single-particle nonergodicity parameter
$f^\text{s}_q$ remains continuous at the transition, which is only
marked by an abrupt but finite change of slope visible in
Fig.~\ref{figscen3fs}. Finally, here as well, the time evolution of
the mean-squared displacement $\delta r^2(t)$, reported in
Fig.~\ref{figscen3msd}, is directly influenced by this anomalous
relaxation of the density fluctuations and, as a result, shows a
regime of continuous critical dynamics in the localized phase near the
glass transition point.

Like in Ref.~\cite{Kra07PRE}, we shall leave aside the peculiar
dynamics found in the vicinity of the state points corresponding to
higher-order singularities or crossing points, either fully embedded
in the liquid-glass transition lines (points E and C) or at their
junctions with the diffusion-localization transition lines (points
T). They certainly are interesting, but such a study would be quite
technical, requiring fine-tuning of the parameters of the models and
refined mathematical tools
\cite{LesHouches,GotSpe02PRE,Spe03PRE,Spe04PRE,GotSpe04JPCM}. So we
prefer to rather concentrate on the generic scenarios, which are the
most likely to be observed in experiments or computer simulations, and
we keep such an advanced discussion for future work, when it will be
useful, for instance, if relevant data become available.

\section{Discussion and conclusion}

In this paper, a MCT for the slow dynamics of a tagged particle moving
in a fluid adsorbed in a disordered porous solid has been
developed. It complements previous work on the collective dynamics
reported in Refs.~\cite{Kra05PRL,Kra05JPCM,Kra07PRE}, so that one
might now consider as complete the mode-coupling theoretical framework
for the study of the dynamics of density fluctuations in QA systems.

From a formal point of view, the newly derived equations keep the
features of those of Ref.~\cite{Kra07PRE} which were found
appealing. They are universality, in the sense that the eventual
dynamical equations do not contain any explicit reference to the
precise nature of the random environment in which the fluid evolves,
and a very close similarity with other mode-coupling equations
previously derived in comparable contexts. So, the overall structure
of the theory appears quite satisfactory.  Note however that this
turns out not to be a completely trivial task to obtain such equations
which display these features and also avoid certain subtle
inconsistencies. Indeed, as shown in Sec.~III, special care seems to
be generically needed in order to properly handle the peculiar
correlations in QA systems, for reasons which are not fully clear at
the moment. In view of these difficulties, it is hoped that the
present approach can be useful in the future as a guide for
developments based on other formalisms.

From a physical point of view, the main result of the present work is
the prediction of a continuous diffusion-localization transition which
can occur before and independently of the liquid-glass transition in
situations of strong confinement. It is characterized by the emergence
of various dynamical anomalies, in particular, by the development of
an algebraic long time tail in the tagged-particle density correlators
and by the opening of a subdiffusive regime in the mean-squared
displacement.

Dealing with the problem of a tagged particle moving in a random
medium, the possibility of a diffusion-localization transition is a
natural expectation. However, the prediction that the single-particle
dynamics can become singular and nonergodic while the collective
dynamics remains regular and ergodic might look suspect, owing to the
fact that the former contributes to the latter, of which it represents
the so-called self part. In fact, the difficulty is only
superficial. Firstly, it should be remembered that the theory for the
collective dynamics and the definition of the ideal liquid-glass
transition are formulated in terms of the \emph{connected} fluid
density correlation functions, while the single-particle density
correlation function is the self part of the \emph{total} fluid
density correlation function. Because the random external field
provided by the fixed matrix induces nonzero average density
fluctuations in the fluid at equilibrium, the latter, defined as
\begin{equation}
\phi^\text{T}_q(t)= \frac{\overline{ \langle
\rho^\text{f}_\mathbf{q}(t) \rho^\text{f}_\mathbf{-q}(0)
\rangle}}{N_\text{f} S^\text{ff}_q}
\end{equation}
with $S^\text{ff}_q = \overline{ \langle \rho^\text{f}_\mathbf{q}
\rho^\text{f}_\mathbf{-q} \rangle} / N_\text{f} = S^\text{ff(c)}_q +
S^\text{ff(b)}_q$, can be written as
\begin{equation}
\phi^\text{T}_q(t) = \frac{S^\text{ff(c)}_q}{S^\text{ff}_q} \phi_q(t)
+ \frac{S^\text{ff(b)}_q}{S^\text{ff}_q},
\end{equation}
from which it is clear that $\phi^\text{T}_q(t)$ never decays to zero
in a QA system. So, one might say that the fluid density fluctuations
are never really ergodic in a QA mixture, in the loose sense that they
saturate at a finite value, like in a bulk glassy system.  Secondly,
based on the results of the MCT for one-component bulk systems
\cite{LesHouches} and on the prevalence of rudimentary approximations
like Vineyard's \cite{Vin58PR}, it is often believed that the
collective and self density fluctuations should generically behave
more or less the same way. This is not correct. In fact, one can
easily demonstrate, using the memory function formalism, that the
total and single-particle density correlation functions are completely
independent, formally decoupled objects \cite{BoonYip}, so that the
general case is rather the contrary, i.e., the two functions display
quite different behaviors \cite{decoupling}. Thus, despite
appearances, the pure diffusion-localization transition scenario
predicted in this work, which is analogous to the one predicted years
ago for bulk binary mixtures with large size asymmetry
\cite{BosTha87PRL,ThaBos91PRA,ThaBos91PRA2}, is not marred by any
inconsistency.

An interesting aspect of this prediction, that we expect to be quite
general, is that it demonstrates the possibility, when strong static
fluid correlations [encoded here in $h^\text{ss(b)}(r)$] are present
in confinement, that the single-particle dynamics develops slow
relaxational features which have no relation whatsoever with glassy
dynamics, if the latter is to be understood in the usual way as a
collective dynamical phenomenon. We believe that this finding could be
important for the design and the interpretation of experiments and
computer simulations on the dynamics of confined glassforming
systems. Indeed, the idea that confinement can be responsible for the
appearence of additional slow dynamical processes which blend with the
glassy dynamics and might obscure its features is definitely not novel
\cite{AlcMcK05JPCM,AlbCoaDosDudGubRadSli06JPCM} (see
Refs.~\cite{ArnStaGorKre96PRE,ArnStaGroHemKre97PRL} for a specific
experimental observation of such a process). But it does not seem to
have been realized previously that there could be such a major
difference in the way some of these processes are reflected in
collective and single-particle quantities, being essentially invisible
in the former and possibly dominant in the latter.  This suggests
that, if one is mainly interested in understanding the effect of
confinement on the glass transition, it should be easier to
concentrate on collective dynamical quantities, and that, if
single-particle data are to be considered, one should be prepared to
face more complex dynamical scenarios which might result from the
interplay between the glassy dynamics and confinement-induced
single-particle relaxation processes.

Finally, all the predictions made in the present and previous papers
\cite{Kra05PRL,Kra05JPCM,Kra07PRE} can in principle be tested by
computer simulations, in order to judge the quality of the proposed
theory. As far as the single-particle dynamics is concerned, this
should not be too challenging, since, in the past few years, a
significant number of simulation studies have actually been already
reported, which precisely dealt with the tagged-particle dynamics in
QA systems
\cite{VirAraMed95PRL,VirMedAra95PRE,GalPelRov02EL,GalPelRov03PRE1,%
GalPelRov03PRE2,AttGalRov05JCP,Kim03EL,ChaJagYet04PRE,MitErrTru06PRE,%
ChaJuaMed08PRE,SunYet08JCP}. Unfortunately, in all these studies,
either the model was simple, but the focus was not on slow dynamics,
or the glassy dynamics was explored, but the fluid model was quite
complex, so that direct comparisons with the results of the present
work are never possible. We hope that the situation can improve very
soon.

\appendix
\section{Replica Ornstein-Zernike equations}
\label{app.oz}
For reference, we quote in this appendix the replica Ornstein-Zernike
equations relating the various pair correlation functions mentioned in
the main text.

For the fluid-matrix part, they read
\cite{GivSte92JCP,LomGivSteWeiLev93PRE,GivSte94PA,RosTarSte94JCP}
\begin{subequations}
\begin{gather}
\hat{h}^\text{mm}_q = \hat{c}^\text{mm}_q + n_\text{m}
\hat{c}^\text{mm}_q \hat{h}^\text{mm}_q, \\
\hat{h}^\text{fm}_q = \hat{c}^\text{fm}_q + n_\text{m}
\hat{c}^\text{fm}_q \hat{h}^\text{mm}_q + n_\text{f}
\hat{c}^\text{ff(c)}_q \hat{h}^\text{fm}_q, \\
\hat{h}^\text{ff(b)}_q = \hat{c}^\text{ff(b)}_q + n_\text{m}
\hat{c}^\text{fm}_q \hat{h}^\text{fm}_q + n_\text{f}
\hat{c}^\text{ff(c)}_q \hat{h}^\text{ff(b)}_q + n_\text{f}
\hat{c}^\text{ff(b)}_q \hat{h}^\text{ff(c)}_q, \\
\hat{h}^\text{ff(c)}_q = \hat{c}^\text{ff(c)}_q + n_\text{f}
\hat{c}^\text{ff(c)}_q \hat{h}^\text{ff(c)}_q,
\end{gather}
\end{subequations}
with $\hat{c}^\text{ff}_q = \hat{c}^\text{ff(c)}_q +
\hat{c}^\text{ff(b)}_q$ and $\hat{h}^\text{ff}_q =
\hat{h}^\text{ff(c)}_q + \hat{h}^\text{ff(b)}_q$. As usual, $h$ and
$c$ denote total and direct correlation functions,
respectively. $\hat{f}$ denotes the Fourier transform of $f$ and the
superscripts have the same meaning as in the main text. 

They can be easily combined with the definitions of the structure
factors,
\begin{subequations}
\begin{gather}
S^\text{mm}_q = 1 + n_\text{m} \hat{h}^\text{mm}_q, \\
S^\text{fm}_q = \sqrt{n_\text{f} n_\text{m}} \hat{h}^\text{fm}_q, \\
S^\text{ff(c)}_q = 1 + n_\text{f} \hat{h}^\text{ff(c)}_q, \\
S^\text{ff(b)}_q = n_\text{f} \hat{h}^\text{ff(b)}_q,
\end{gather}
\end{subequations}
(remember that $S^\text{ff}_q=S^\text{ff(c)}_q + S^\text{ff(b)}_q$,
hence $S^\text{ff}_q= 1 + n_\text{f} \hat{h}^\text{ff}_q$) to express
them in terms of the direct correlation functions (see Appendix A of
Ref.~\cite{Kra07PRE}).

For the single-particle part, the relevant equations, which are
limiting cases of those reported in Ref.~\cite{PasKah00PRE}, read
\begin{subequations}
\begin{gather}
\hat{h}^\text{sm}_{q} = \hat{c}^\text{sm}_{q} + n_\text{m}
\hat{c}^\text{sm}_{q} \hat{h}^\text{mm}_{q} + n_\text{f}
\hat{c}^\text{sf(c)}_{q} \hat{h}^\text{fm}_{q}, \\
\hat{h}^\text{sf(b)}_{q} = \hat{c}^\text{sf(b)}_{q} + n_\text{m}
\hat{c}^\text{sm}_{q} \hat{h}^\text{fm}_{q} + n_\text{f}
\hat{c}^\text{sf(b)}_{q} \hat{h}^\text{ff(c)}_{q} + n_\text{f}
\hat{c}^\text{sf(c)}_{q} \hat{h}^\text{ff(b)}_{q}, \\
\hat{h}^\text{sf(c)}_{q} = \hat{c}^\text{sf(c)}_{q} + n_\text{f}
\hat{c}^\text{sf(c)}_{q} \hat{h}^\text{ff(c)}_{q}, \\
\hat{h}^\text{ss(b)}_{q} = \hat{c}^\text{ss(b)}_{q} + n_\text{m}
\hat{c}^\text{sm}_{q} \hat{h}^\text{sm}_{q} + n_\text{f}
\hat{c}^\text{sf(c)}_{q} \hat{h}^\text{sf(b)}_{q} + n_\text{f}
\hat{c}^\text{sf(b)}_{q} \hat{h}^\text{sf(c)}_{q}, \\
\hat{h}^\text{ss(c)}_{q} = \hat{c}^\text{ss(c)}_{q} + n_\text{f}
\hat{c}^\text{sf(c)}_{q} \hat{h}^\text{sf(c)}_{q},
\end{gather}
\end{subequations}
with $\hat{c}^\text{sf}_{q} = \hat{c}^\text{sf(c)}_{q} +
\hat{c}^\text{sf(b)}_{q}$, $\hat{h}^\text{sf}_{q} =
\hat{h}^\text{sf(c)}_{q} + \hat{h}^\text{sf(b)}_{q}$,
$\hat{c}^\text{ss}_{q} = \hat{c}^\text{ss(c)}_{q} +
\hat{c}^\text{ss(b)}_{q}$, and $\hat{h}^\text{ss}_{q} =
\hat{h}^\text{ss(c)}_{q} + \hat{h}^\text{ss(b)}_{q}$.

\section{Mode-coupling equations for a mixture adsorbed in a
  disordered porous solid}
\label{MCTmixture}

In this appendix, we report the MCT equations for the dynamics of a
mixture adsorbed in a disordered porous matrix. They are derived
following an obvious generalization of the procedure outlined in
appendix B of Ref.~\cite{Kra07PRE} and are thus given without proof.

The motivation for this appendix is twofold. First, these equations
are needed in order to derive those for the tagged-particle dynamics,
which are the main result of the present work. Second, most simple
glassformer models which are considered in the literature for
comparisons between theoretical predictions and intensive computer
simulation results are binary mixtures, for the reason that they are
much less prone to crystallisation than one-component systems.  So it
is likely that sooner or later a version of the theory dealing with
adsorbed mixtures will become useful, for instance, in order to
perform quantitative tests like the ones reported in
Refs.~\cite{NauKob97PRE,KobNauSci02JNCS} for bulk systems.

We consider a quenched-annealed multicomponent mixture containing $n$
fluid components. Each of them consists of $N_\alpha$ particles of
mass $m_\alpha$ ($1\leq\alpha\leq n$). The total fluid particle number
is $N_\text{f}=\sum_{\alpha=1}^{n} N_\alpha$ and the fluid number
fractions are defined as $x_\alpha=N_\alpha/N_\text{f}$. The system is
at temperature $T$ and has volume $V$, hence the total fluid density
is $n_\text{f}=N_\text{f}/V$. Note that it is not necessary to
provide any information on the disordered matrix, since only
quantities characterizing the fluid component of the QA system will
appear in the final results.

The dynamical variables of interest are the time-dependent density
fluctuations of the different fluid species,
\begin{equation}
\rho^\alpha_\mathbf{q}(t) = \sum_{j=1}^{N_\alpha} e^{i \mathbf{q}
\mathbf{r}^\alpha_j(t)},
\end{equation}
where $\mathbf{r}^\alpha_j(t)$ is the position of the fluid particle
$j$ of type $\alpha$ at time $t$. From these quantities, one can form
the relevant static structure factors, connected,
\begin{equation}
S^{\alpha\beta\text{(c)}}_q = \frac1{N_\text{f}} \overline{\langle
\delta\rho^\alpha_\mathbf{q} \delta\rho^\beta_\mathbf{-q}
\rangle},
\end{equation}
with $\delta\rho^\alpha_\mathbf{q}(t) = \rho^\alpha_\mathbf{q}(t) -
\langle \rho^\alpha_\mathbf{q} \rangle$, and blocked or disconnected,
\begin{equation}
S^{\alpha\beta\text{(b)}}_q = \frac1{N_\text{f}} \overline{\langle
\rho^\alpha_\mathbf{q} \rangle \langle \rho^\beta_\mathbf{-q} \rangle},
\end{equation}
as well as the nonnormalized time-dependent connected density
fluctuation autocorrelation functions
\begin{equation}
F^{\alpha\beta}_q(t) = \frac1{N_\text{f}} \overline{\langle
\delta\rho^\alpha_\mathbf{q}(t) \delta\rho^\beta_\mathbf{-q}(0)
\rangle}.
\end{equation}

These functions obey standard generalized Langevin equations which
read in matrix form,
\begin{equation}
\ddot{\mathbf{F}}_q(t)+\mathbf{\Omega}^2_q
\mathbf{F}_q(t)+\int_0^t d\tau \mathbf{M}_q(t-\tau)
\dot{\mathbf{F}}_q(\tau)=\mathbf{0},
\end{equation}
with initial conditions $\mathbf{F}_q(0)=\mathbf{S}^\text{(c)}_q$ and
$\dot{\mathbf{F}}_q(0)=\mathbf{0}$, and a frequency matrix given by
\begin{equation} 
[\mathbf{\Omega}^2_q]^{\alpha\beta}= \frac{q^2 k_B T
x_\alpha}{m_\alpha} [\mathbf{S}^{\text{(c)}-1}_q]^{\alpha\beta},
\end{equation}
where $\mathbf{S}^{\text{(c)}-1}_q$ is the matrix inverse of
$\mathbf{S}^\text{(c)}_q$.

The MCT provides an expression for the kernel $\mathbf{M}_q(t)$, which
reads $\mathbf{M}_q(t)=\mathbf{\Gamma}_q
\delta(t)+\mathbf{M}^{(\text{MC})}_q(t)$, where $\mathbf{\Gamma}_q$ is
a matrix of friction coefficients associated with fast dynamical
processes and
\begin{equation}
M^{\alpha\alpha'(\text{MC})}_q(t) = \frac{q^2 k_B T}{m_\alpha
x_{\alpha'}} \left[ m^{\alpha\alpha'(2)}_q(t) +
m^{\alpha\alpha'(1)}_q(t) \right],
\end{equation}
with 
\begin{equation}
m^{\alpha\alpha'(2)}_q(t) = \frac12 n_\text{f} \int
\frac{d^3\mathbf{k}}{(2\pi)^3}
\sum_{\beta,\gamma,\beta',\gamma'=1}^{n}
V^{\alpha\beta\gamma(2)}_\mathbf{q,k}
V^{\alpha'\beta'\gamma'(2)}_\mathbf{q,k} F^{\beta\beta'}_k(t)
F^{\gamma\gamma'}_{|\mathbf{q-k}|}(t)
\end{equation}
and
\begin{equation}
m^{\alpha\alpha'(1)}_q(t) = n_\text{f} \int
\frac{d^3\mathbf{k}}{(2\pi)^3} \sum_{\beta,\beta'=1}^{n}
V^{\alpha\beta(1)}_\mathbf{q,k} V^{\alpha'\beta'(1)}_\mathbf{q,k}
F^{\beta\beta'}_k(t) S^{\alpha\alpha'\text{(b)}}_{|\mathbf{q-k}|}.
\end{equation}
The vertices are given by
\begin{align}
V^{\alpha\beta\gamma(2)}_\mathbf{q,k} 
= & \frac{\mathbf{q}\cdot\mathbf{k}}{q^2} \delta_{\alpha\gamma}
\hat{c}^{\alpha\beta\text{(c)}}_k +
\frac{\mathbf{q}\cdot(\mathbf{q-k})}{q^2} \delta_{\alpha\beta}
\hat{c}^{\alpha\gamma\text{(c)}}_{|\mathbf{q-k}|}, \\
V^{\alpha\beta(1)}_\mathbf{q,k} 
= & \frac{\mathbf{q}\cdot\mathbf{k}}{q^2}
\hat{c}^{\alpha\beta\text{(c)}}_k +
\frac{\mathbf{q}\cdot\mathbf{(q-k)}}{q^2} \frac{1}{n_\text{f}
x_\alpha} \delta_{\alpha\beta},
\end{align}
where the $\hat{c}^{\alpha\beta\text{(c)}}_q$'s are the Fourier
transforms of the connected direct correlation functions
\cite{PasKah00PRE}.

The equations for the tagged-particle dynamics are obtained by
considering a species $\sigma$ in the limit of a vanishing number
fraction. One then has $\phi^\text{s}_q(t) = \lim_{x_\sigma \to 0}
F^{\sigma\sigma}_q(t) / x_\sigma$. After tedious but straightforward
algebra, one finds the usual Langevin equation \eqref{eqlangevinself}
and
\begin{equation}
m^{\text{s(MC)}}_q(t) = m^{\text{s}(2)}_q(t) + m^{\text{s}(1)}_q(t),
\end{equation}
with
\begin{equation}
m^{\text{s}(2)}_q(t) = n_\text{f} \int \frac{d^3\mathbf{k}}{(2\pi)^3}
\sum_{\beta,\beta'=1}^n \left[
\frac{\mathbf{q}\cdot(\mathbf{q-k})}{q^2} \right]^2
\hat{c}^{\text{s}\beta\text{(c)}}_{|\mathbf{q-k}|}
\hat{c}^{\text{s}\beta'\text{(c)}}_{|\mathbf{q-k}|}
F^{\beta\beta'}_{|\mathbf{q-k}|}(t) \phi^\text{s}_k(t)
\end{equation}
and
\begin{equation}
m^{\text{s}(1)}_q(t) = \int \frac{d^3\mathbf{k}}{(2\pi)^3} \left[
\frac{\mathbf{q}\cdot(\mathbf{q-k})}{q^2} \right]^2
h^\text{ss(b)}_{|\mathbf{q-k}|} \phi^\text{s}_k(t).
\end{equation}
The $\hat{c}^\text{s$\alpha$(c)}_{q}$'s and $\hat{h}^\text{ss(b)}_{q}$
are the Fourier transforms of the connected single-particle-fluid
direct correlation functions and of the
single-particle-single-particle blocked total correlation function,
respectively \cite{PasKah00PRE}.

The equations reported in the main body of the paper are immediately
obtained for $n=1$.

\section{Hydrodynamic approximation for the diffusion-localization
  transition}
\label{hydro}

In this appendix, the results of a generalized hydrodynamic
approximation, in which the wave vector dependence of the MCT
equations is simplified, are reported for the dynamics in the vicinity
of the diffusion-localization transition. They represent an extension
of Leutheusser's theory \cite{Leu83PRA} to finite fluid densities.

The theory is best obtained by considering Laplace transformed
dynamical quantities, defined as $\tilde{\phi}(z)=i\int_0^\infty dt
\,e^{izt} \phi(t)$ for $\text{Im } z > 0$. Equation
\eqref{eqlangevinself} then becomes
\begin{equation}\label{gendiff}
\tilde{\phi}^\text{s}_q(z)= \frac{-1}{z+q^2 \tilde{D}_q(z)}
\end{equation}
with the generalized diffusion coefficient $\tilde{D}_q(z)$ given by
\begin{equation}
\tilde{D}_q(z) = \frac{- k_B T / m_\text{s}}{z + \omega_{q}^2
\tilde{m}^\text{s}_q(z)}.
\end{equation}
The hydrodynamic approximation simply amounts to neglecting the $q$
dependence of $\tilde{D}_q(z)$ in Eq.~\eqref{gendiff}, where it is
replaced by its $q\to0$ limit
\begin{equation}\label{hydroapp}
\tilde{D}(z) = \frac{- k_B T / m_\text{s}}{z +
\tilde{m}^\text{MSD}(z) k_B T / m_\text{s} }.
\end{equation}

Further progress is made by noting that it is not necessary to give a
detailed account of the bilinear mode-coupling term in
$m^\text{MSD}(t)$ if one is interested in the anomalous dynamics near
the diffusion-localization threshold. Indeed, on the one hand, it is
clear that this term can be singular when $n_\text{f}$ is very small,
because the diffusion-localization and liquid-glass transition lines
are then very close, but it is also very small, because of the
explicit factor $n_\text{f}$ in $w^{(2)}_{k}$. In particular, it
vanishes identically when $n_\text{f}=0$. It can thus be neglected for
small $n_\text{f}$. On the other hand, for larger $n_\text{f}$, but
not near point T which would require a refined treatment, the wide
separation between the two transition lines guarantees that the
collective dynamics will be fast and regular at the
diffusion-localization transition. It results that the singular
tagged-particle dynamics is cut off in the bilinear term, which
generates a fast and smoothly varying contribution to
$m^\text{MSD}(t)$. Thus it is in general enough to set
\begin{equation}\label{hydrokernel}
m^\text{MSD}(t) = \gamma \delta(t) + \int_0^{+\infty} \frac{k^4 dk}{6
\pi^2} \hat{h}^\text{ss(b)}_{k} \phi^\text{s}_{k}(t)
\end{equation}
with an appropriate renormalization of $\gamma$.  Then, combining
Eqs.~\eqref{gendiff} with $\tilde{D}_q(z) \simeq \tilde{D}(z)$,
\eqref{hydroapp}, and the Laplace transform of
Eq.~\eqref{hydrokernel}, a transcendental equation for $\tilde{D}(z)$
is obtained, which reads
\begin{equation}\label{transcendental}
\left(\frac{m_\text{s}}{k_B T} z + i \gamma \right) \tilde{D}(z) = - 1
+ \int_0^{+\infty} \frac{k^4 dk}{6 \pi^2}
\frac{\hat{h}^\text{ss(b)}_{k}}{k^2+z/\tilde{D}(z)}.
\end{equation}
From the solution of this equation, which is the core result of the
proposed generalized hydrodynamic approximation, all relevant
dynamical quantities can be computed.

The diffusive and localized phases are characterized by different
small $z$ behaviors of $\tilde{D}(z)$. In the diffusive phase,
\begin{equation}
\lim_{z\to i0} \tilde{D}(z) = i D,
\end{equation}
where $D$ is the finite long-time diffusion coefficient. From
Eq.~\eqref{transcendental}, one finds
\begin{equation}
D = \frac{1}{\gamma} \left[ 1 - \frac13 h^\text{ss(b)}(r=0) \right].
\end{equation}
In the localized phase, 
\begin{equation}
\tilde{D}(z) \sim z r_\text{l}^2,
\end{equation}
where $r_\text{l}$ is the finite localization length, which obeys
\begin{equation} \label{hydrolength}
\int_0^{+\infty} \frac{k^4 dk}{6 \pi^2}
\frac{\hat{h}^\text{ss(b)}_{k}}{k^2+r_\text{l}^{-2}} = 1.
\end{equation}
Using a large $r_\text{l}$ expansion of Eq.~\eqref{hydrolength}, one
easily shows that the divergence of $r_\text{l}$ and the vanishing of
$D$ both occur when
\begin{equation}
h^\text{ss(b)}(r=0) = 3.
\end{equation}
So, as anticipated, the present theory indeed describes a \emph{bona
fide} diffusion-localization transition scenario, with the above
condition providing an implicit equation of the transition line.

\end{document}